\documentclass[showpacs,preprintnumbers,amsmath,amssymb,twocolumn,floatfix]{revtex4-1}

\usepackage{amsmath}
\usepackage{graphicx}
\usepackage{epsf}
\usepackage{psfrag}
\usepackage{epsfig}
\usepackage{graphics}
\usepackage{pstricks}
\usepackage{pst-node}
\usepackage{enumerate}

\newcommand{\sq}[1]{\ensuremath{\mathsf{#1}}}

\newcommand{\pw}[1]{\psframebox[linewidth=0.4pt]{\sq{#1}}}

\psset{linearc=0.15}

\begin{document}

\pagenumbering{arabic}

\title{{\bf Systematic Implementation of Implicit Regularization for Multi-Loop 
Feynman Diagrams}}

\newcommand{\be}{\begin{equation}}
\newcommand{\ee}{\end{equation}}

\date{\today}

\author{A. L. Cherchiglia$^{(a)}$} \email []{adriano@fisica.ufmg.br}
\author{Marcos Sampaio$^{(a)}$}\email[]{msampaio@fisica.ufmg.br}
\author{M. C. Nemes$^{(a)}$}\email[]{carolina@fisica.ufmg.br}

\affiliation{(a) Federal University of Minas Gerais - Physics
Department - ICEx \\ P.O. BOX 702, 30.161-970, Belo Horizonte MG -
Brazil}
\begin{abstract}

\noindent
Implicit Regularization (IReg) is a candidate to become an invariant framework in momentum space to perform Feynman diagram calculations to arbitrary loop order. In this work we present a systematic implementation of our method that automatically displays the terms to be subtracted by Bogoliubov's recursion formula. Therefore, we achieve a twofold objective: we show that the IReg program respects unitarity, locality and Lorentz invariance and we show that our method is consistent since we are able to display the divergent content of a multi-loop amplitude in a well defined set of basic divergent integrals in one loop momentum only which is the essence of IReg. Moreover, we conjecture that momentum routing invariance in the loops, which has been shown to be connected with gauge symmetry, is a fundamental symmetry of any Feynman diagram in a renormalizable quantum field theory. 
\end{abstract}


\maketitle

\section{Introduction}

A consistent renormalization program in QFT appeared after the work of Bogoliubov, Parasiuk, Hepp and Zimmerman (BPHZ) \cite{Bogoliubov:1957gp}-\cite{Zimmerman2} in which a prescription to extract recursively the divergences of a multi-loop Feynman graph complying with unitarity, locality and Lorentz invariance was presented. The BPHZ program generalized the Dyson's subtraction to general overlapping diagrams to arbitrary loop order leading to the concept of renormalizable quantum field theory. Such program systematizes, according to the topology of the graph, the subtraction necessary to render the corresponding amplitude finite through the forest formula. The proof of finitude provided by the latter is by construction regularization independent. However, for concrete predictions such as scattering amplitudes in collision processes of elementary particles, the method of Dimensional Regularization (DReg) and minimal subtraction \cite{Bollini:1972ui}, \cite{'tHooft:1972fi} combined with Zimmerman's forest formula has proven to be an efficient and successful calculational tool particularly for gauge theories. The forest formula can be casted into a counterterm language by means of Bogoliubov's recursion formula \cite{Muta}, complying with locality, Lorentz invariance, unitarity and causality.

To calculate S-matrix elements in a symmetry preserving fashion in a quantum field theoretical model sensitive to dimensional continuation on the space-time, the problem is more subtle. 
The construction of an invariant regularization framework is aesthetically more appealing but this is not the main motivation. Although in one hand imposing constraint equations derived from Ward identities order by order in perturbation theory obliterates the need of an invariant regularization, on the other hand it renders the calculation more involved from the calculational viewpoint. Besides, if quantum symmetry breakings occur in perturbation theory, an invariant scheme is essential to judge it as physical or spurious. Supersymmetric gauge theories are conspicuous examples of models in which regularization and renormalization play a fundamental role especially as new accurate experimental evidence, viz. electroweak precision observables \cite{Eletroweak}-\cite{Eletroweak4}, demands consistent theoretical calculations higher than one loop order to understand physics beyond the Standard Model.

Therefore, the construction of an invariant regularization is justified and, in order to be as reliable as DReg (wherever DReg can be applied), it must be shown to comply with locality, Lorentz invariance, unitarity and causality. Recently, an invariant regularization framework (IReg) has been developed and shown to be consistent and symmetry preserving in several instances \cite{Battistel:1998sz}-\cite{QED}. The essence of the method is to write the divergences in terms of loop integrals in one internal momentum which do not need to be explicitly evaluated. Moreover it acts in the physical dimension of the theory and gauge invariance is controlled by regularization dependent surface terms which when set to zero define a constrained version of IReg (CIReg) and deliver gauge invariant amplitudes automatically. Therefore it is in principle applicable to all physical relevant quantum field theories, supersymmetric gauge theories included. A non trivial question is whether we can generalize this program to arbitrary loop order in consonance with locality, unitarity and Lorentz invariance, especially when overlapping divergences occur. This is the main subject of this work in which we use the simplest renormalizable field theoretical model to show how to implement IReg in such a way that it displays all the terms to be subtracted by Bogoliubov's recursion formula automatically. All the other physical theories can be treated within the same strategy after space time and internal algebra are performed. Another result of this contribution is to show that if the surface terms are not set to zero they will contaminate the renormalization group coefficients. Thus, we are forced to adopt CIReg which is equivalent to demand momentum routing invariance in the loops. This feature leads us to conjecture that momentum routing invariance is a fundamental symmetry of any Feynman diagram.  

\section{The rules of implicit regularization}
\label{s:overview}

We restrict ourselves to massless theories and power counting infrared safe integrals. The first restriction is justified because, as we show in Section \ref{s:mass}, to implement a mass independent renormalization scheme in IReg we need only the massless basic divergent integrals that we present below. When infrared divergences do appear, a dual version of IReg operating in coordinate space displays infrared divergences as basic divergent integrals as well, in a way that infrared and ultraviolet degrees of freedom are clearly distinguished \cite{Infra}-\cite{Infrared2}.

Given the amplitude of a $n$-loop Feynman graph with $L$ external legs, the basic strategy of IReg is to free all divergences of external momenta and express them in terms of basic divergent integrals in one loop momentum only. To achieve this purpose, we need to perform $(n-1)$ integrations, but the order in which they are performed is not clear a priori. In the next section we present a systematic way to choose the order of integration which, as a byproduct, displays the counterterms to be subtracted by Bogoliubov's recursion formula. Considering that we made this choice, we can redefine the internal momenta in such a way that the integral in $k_{l}$ is the l-th we are going to deal with and it is typically of the form
\begin{align}
&I^{\nu_{1}\ldots \nu_{m}}\!=\!\!\int\limits_{k_{l}}\!\frac{A^{\nu_{1}\ldots \nu_{m}}(k_{l},q_{i})}{\prod_{i}[(k_{l}-q_{i})^{2}-\mu^{2}]}\ln^{l-1}\!\left(\!-\frac{k_{l}^{2}-\mu^{2}}{\lambda^{2}}\right)\!,
\label{I}
\end{align}
\noindent
where $l=1\cdots n$. In the above equation, $q_{i}$ is an element (or combination of elements) of the set $\{p_{1},\ldots,p_{L},k_{l+1},\ldots,k_{n}\}$, $\int_{k_{l}}\equiv\int d^dk_{l}/(2 \pi)^d$ and $\mu^{2}$ is an infrared regulator. 

Since the original integral is infrared safe, the limit $\mu^2\rightarrow 0$ is well-defined and must be taken in the end of the calculation. The logarithmical dependence appears because this is the characteristic behaviour of the finite part of massless amplitudes \cite{Delamotte}. $\lambda$ is an arbitrary non-vanishing parameter with dimension of mass which parametrizes the freedom one has to subtract the divergences (renormalization group scale). It appears at one loop level and survives to higher orders through a regularization independent mathematical identity (eq. \ref{scale}) as we show in the end of this section. The function $A^{\nu_{1}\ldots \nu_{m}}(k_{l},q_{i})$ may contain constants and all possible combinations of $k_{l}$ and $q_{i}$ compatible with the Lorentz structure. Care must be exercised when it contains a term like $(k_{l}-q_{i})^{2}$. In this case, we must cancel it against one of the denominators because, as we are dealing with divergent integrals, symmetric integration is a forbidden operation \cite{CarlosDfR}, \cite{PerezVictoria:2001ej}.

Now, we apply the rules of IReg. Assuming that a regulator $\Lambda$ is implicit in the integral, we can use the following mathematical identity in the denominators:
\begin{align}
&\frac{1}{(k_{l}-q_{i})^2-\mu^2}\;=\;\sum_{j=0}^{n_{i}^{(k_{l})}-1}\frac{(-1)^{j}(q_{i}^2-2q_{i} \cdot k_{l})^{j}}{(k_{l}^2-\mu^2)^{j+1}}\quad\nonumber \\
&\quad\quad\quad\quad\;+\frac{(-1)^{n_{i}^{(k_{l})}}(q_{i}^2-2q_{i} \cdot k_{l})^{n_{i}^{(k_{l})}}}{(k_{l}^2-\mu^2)^{n_{i}^{(k_{l})}}
\left[(k_{l}-q_{i})^2-\mu^2\right]}. 
\label{ident}
\end{align}

The values of $n_{i}^{(k_{l})}$ are chosen such that all divergent integrals have a denominator free of $q_{i}$.

After the use of (\ref{ident}), the divergent integrals can be casted as a combination of
\begin{align}
I_{log}^{(l)}(\mu^2)\equiv \!\!\int\limits_{k_{l}}^{\Lambda} \!\frac{1}{(k_{l}^2\!-\mu^2)^{d/2}}
\ln^{l-1}\!{\left(\!\!-\frac{k_{l}^2\!-\mu^2}{\lambda^2}\right)}\!,
\end{align}
\noindent
and
\begin{align}
I_{log}^{(l)\nu_{1} \cdots \nu_{r}}(\mu^2)\equiv \!\!\int\limits_{k_{l}}^{\Lambda} \!\frac{k_{l}^{\nu_1}\!\cdots
k_{l}^{\nu_{r}}}{(k_{l}^2\!-\mu^2)^\beta}
\ln^{l-1}\!{\left(\!\!-\frac{k_{l}^2\!-\mu^2}{\lambda^2}\right)}\!.
\label{IlogLorentz}
\end{align}

In the above formula, the subscript $log$ means that we are dealing with a logarithmic divergent integral ($r=2\beta-d$). It is important to note that only this type of divergence appears because linear and quadratic divergent integrals vanish for massless theories \cite{Carlos}, \cite{Note1}.

Although we have already reduced the divergences to basic divergent integrals free of external momenta, we can show that the integrals defined above are related. For example, in a case in which $r=2$ we have
\begin{align}
&I_{log}^{(l)\,\mu 
\nu}(\mu^2)=\sum_{j=1}^{l}\left(\frac{2}{d}\right)^j\frac{(l-1)!}{(l-j)!}\Bigg\{\frac{g^{\mu \nu}}{2}I_{log}^{(l-j+1)}(\mu^2)\nonumber\\
&-\frac{g^{\mu \rho}}{2}\int\limits_{k}\frac{\partial}{\partial k^{\rho}}\Bigg[\frac{k^{\nu}}{(k^2-\mu^2)^{d/2}}\ln^{l-j}{\left(-\frac{k^2\!-\mu^2}{\lambda^2}\right)}\Bigg]\!\Bigg\},\nonumber\\
\end{align}
\noindent
or equivalently, for short
\begin{align}
&I_{log}^{(l)\,\mu 
\nu}(\mu^2)-g^{\mu \nu}\sum\limits_{j=1}^{l}\;a_{j}\;I_{log}^{(l-j+1)}(\mu^2)=\Upsilon_{l}\;g^{\mu \nu},\nonumber\\
&a_{j}\equiv\frac{1}{2}\left(\frac{2}{d}\right)^j\!\frac{(l-1)!}{(l-j)!}.
\label{identsurface}
\end{align}

In the previous equation, $\Upsilon_{l}$ is a surface term which is arbitrary and in general regularization dependent. It was shown in \cite{Battistel:1998sz}-\cite{CarlosDfR} that setting all surfaces terms to zero defines a constrained version of IReg (CIReg) and corresponds to invoking momentum routing invariance in the loops of a Feynman graph. This in turn is related to gauge invariance and it was shown that adopting CIReg is a sufficient condition to ensure gauge symmetry \cite{QED}. One may verify that Dimensional Regularization (DReg) evaluates the surface terms to zero what demonstrates that CIReg and DReg are compatible. In theories with less symmetry content such as scalar field theories one could ask whether momentum routing invariance plays any relevant role. We shall answer this question by calculating the first two coefficients of the $\phi^{3}_{6}$ theory $\beta$ function which are universal. We verify that the arbitrarity introduced by the surface terms cannot be hidden in the redefinition of a renormalization scheme. This feature leads us to conjecture that momentum routing invariance is a fundamental symmetry of Feynman diagrams.

At this point we notice that the divergences can be written in terms of one object namely
\begin{align}
I_{log}^{(l)}(\mu^2)\equiv\! \int\limits_{k_{l}}^{\Lambda}\!\! \frac{1}{(k_{l}^2-\mu^2)^{d/2}}
\ln^{l-1}\!\left(\!-\frac{k_{l}^2-\mu^2}{\lambda^2}\right)\!\!.
\label{Ilog(n)}
\end{align}

However, the above integral is ultraviolet and infrared divergent as $\mu^2\rightarrow 0$. To separate these divergences and define a genuine ultraviolet divergent object we use the scale relation below \cite{Note2}
\begin{align}
&I_{log}^{(l)}(\mu^2)=I_{log}^{(l)}(\lambda^2)-\frac{b_{d}}{l}\ln^{l}\left(\frac{\mu^2}{\lambda^2}\right)+\nonumber\\&\quad\quad  b_{d}\sum_{k=1}^{A}\binom{A}{k}\sum_{j=1}^{l-1}\frac{(-1)^k}{k^j}\frac{(l-1)!}{(l-j)!}\ln^{l-j}\left(\frac{\mu^2}{\lambda^2}\right)\!,
\label{scale}\\
&\lambda^{2}\neq0,\;\; A\equiv\frac{(d-2)}{2} \mbox{,} \;\; b_{d}\equiv\frac{i}{(4\pi)^{d/2}}\frac{(-1)^{d/2}}{\Gamma(d/2)}. 
\end{align}

Since we are dealing with infrared safe models the infrared divergence must disappear in the amplitude as a whole. This in fact occurs because, as we use identity (\ref{ident}), the finite part of the amplitude will also have a logarithmical dependence in $\mu^2$ and it is just the one expected to cancel the infrared divergence coming from the use of the scale relation. As mentioned before, we note that $\lambda$ parametrizes the freedom we have to subtract the divergences and becomes a natural candidate for a renormalization group scale. 

We repeat the above procedure until we are left with only one integral in the internal momentum and, consequently, we are able to express all divergences in terms of $I_{log}^{(n)}(\lambda^2)$. One of the purposes of the next sections is to show, for a general $n$-loop Feynman graph, how this program can be implemented in a systematic way which is compatible with Bogoliubov's recursion formula. 

\section{Systematic implementation of Bogoliubov's recursion formula in IReg}
\label{s:BPHZ}

In this section we develop an algorithm that implements IReg to multi-loop Feynman graphs. It is constructed in such a way that it displays the terms to be subtracted by Bogoliubov's recursion formula and thus fulfilling unitarity, locality and Lorentz invariance. 

In order to implement IReg in a systematic way to a $n$-loop Feynman graph we adapt identity (\ref{ident}), which was initially conceived for one-loop order, to arbitrary order since it does not furnish us a natural sequence in which the integrals must be performed. Therefore, our first task is to rewrite it in such a way that it evinces the divergent behaviour of the amplitude as the internal momenta go to infinity in all possible ways. Restricting $q_{i}$ in (\ref{ident}) to be external momenta and expanding $(p_i^2-2p_i \cdot k)^{j}$ with the familiar binomial formula yields,
\begin{align}
\frac{1}{(k-p_i)^2-\mu^2}=\!\sum_{l=0}^{2(n_{i}^{(k)}-1)}f_{l}^{\;(k,\;p_{i})}+{\bar{f}}^{\;(k,\;p_{i})}
\label{identbphz},
\end{align}
where we defined,
\begin{align}
&f_{l}^{\;(k,\;p_{i})}\equiv\sum_{j=0}^{\left\lfloor l/2\right\rfloor}\Theta(B)\binom{l-j}{j}
\frac{(-p_i^2)^{j}(2p_i \cdot\ k)^{l-2j}}{(k^2-\mu^2)^{l+1-j}},\\
&{\bar{f}}^{\;(k,\;p_{i})}
\equiv\frac{(-1)^{n_{i}^{(k)}}(p_i^2-2p_i \cdot k)^{n_{i}^{(k)}}}{(k^2-\mu^2)^{n_{i}^{(k)}}
\left[(k-p_i)^2-\mu^2\right]},\label{fbar}\\
&\Theta(x)\equiv\left\{\begin{array}{rc}
0&\mbox{if}\quad x\leq 0\\
1 &\mbox{if}\quad x>0
\end{array}\right.\nonumber,\\
&B\equiv n_{i}^{(k)}+j-l,\quad\left\lfloor x \right\rfloor\equiv\mbox{max}\{n\in\mathcal{Z}|n\leq x\}\nonumber.
\end{align}

The terms $f_{l}^{\;(k,\;p_{i})}$ are constructed in such a way that they behave like $k^{-(l+2)}$ as $k\rightarrow \infty$ and we choose the values of $n_{i}^{(k)}$ in order to assure the UV finitude of ${\bar{f}}^{(k,\;p_{i})}$. The above identity is the keystone of our procedure which, when applied to a given Feynman graph, can be summarized in the following steps:

\begin{enumerate}[A.]
\item Identify the propagators which depend on the external momenta of the graph and apply identity (\ref{identbphz});
\label{pro}
\item Find out the minimum value of $n_{j}^{(k_{i})}$ needed to assure the finitude of the terms that contain ${\bar{f}}^{\;(k_{i},\;p_{j})}$ as $k_{i}\rightarrow \infty$ in all possible ways;
\label{n}
\item Repeat the above step for all propagators identified in step \ref{pro};
\label{repeat}
\item Identify the divergent terms and classify them according to all possible ways that the internal momenta approach infinity;
\label{identify}
\item Use the rules of IReg (in the way presented in Section \ref{s:overview}) in the terms identified in step \ref{identify} according to their classification;
\label{apIR}
\item Set aside the divergent terms that contain $I_{log}^{(l)}(\lambda^{2})$ and apply the procedure again on the ones that do not.
\label{save}
\end{enumerate} 

After step \ref{save}, we have only two kind of terms: the ones in which $I_{log}^{(l)}(\lambda^{2})$ multiplies an integral and the ones in which $I_{log}^{(l)}(\lambda^{2})$ multiplies only constants and/or polynomials in the external momenta. The first are just the terms cancelled by Bogoliubov's recursion formula while the latter are the typical divergence of the graph, i.e. after subtraction of subdivergences. In other words, our prescription implements IReg in a way that displays automatically the terms cancelled by Bogoliubov's recursion formula.

We illustrate its applicability with some examples. Since we are concerned only with the structure of the divergences, we work with the simplest renormalizable quantum field theory: massless $\phi_{6}^{3}$. In this theory, only graphs up to three external legs are divergent \cite{Muta}. The graphs with one external leg have only quadratic divergences and these always vanish for massless theories \cite{Carlos}. Therefore, the graphs we deal with have only two or three external legs and correspond to the renormalization of the propagator and the vertex functions respectively. 

In all the examples we present we choose a particular momentum routing in order to simplify our calculation. However, we could have chosen a different set of internal momenta and, if we have done so, we would obtain the same divergent structure expressed as basic divergent integrals. In other words, our prescription is not limited to a specific choice of momentum routing in the internal lines of a Feynman graph.

\subsection{One- and two-loop self-energy and vertex diagrams}

We begin with the one-loop correction for the propagator whose graph is
\begin{figure}[h!]
\begin{center}
\includegraphics{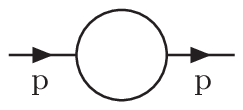}
\end{center}
\caption{Graph $P^{(1)}$}
\label{p1}
\end{figure}

The amplitude depicted by fig. \ref{p1} reads
\begin{align}
\Xi^{(1)}\!\equiv\!\frac{g^{2}}{2}\!\int\limits_{k}\!\frac{1}{(k^{2}-\mu^{2})}\!\frac{1}{[(k-p)^{2}-\mu^2]}\mbox{,}\quad\!\!\!\!\int\limits_{k}\!\!\equiv\!\!\int\!\!\frac{d^6k}{(2 \pi)^6}.
\end{align}

Notice that we have introduced an infrared regulator in the denominators. We use identity (\ref{identbphz}) in the denominator that contains the external momentum to obtain
\begin{align}
\frac{\Xi^{(1)}}{g^{2}}=\frac{1}{2}\!\int\limits_{k}\!\frac{1}{(k^{2}-\mu^{2})}\left[\sum_{l=0}^{2(n^{(k)}-1)}f_{l}^{\;(k,\;p)}+{\bar{f}}^{\;(k,\;p)}\right]\!\!.
\end{align}

We must choose $n^{(k)}$ in order to guarantee the finitude of the term that contains ${\bar{f}}^{\;(k,\;p)}$. By power counting we find that $n^{(k)}>2$ and thus we choose $n^{(k)}=3$. Having found the value of $n^{(k)}$ we can extract the divergent terms. As $f_{l}^{\;(k,\;p)}$ goes like $k^{-(l+2)}$ we find by power counting that they are given by:

\begin{enumerate}
\item Quadratic divergence
\begin{align}
\int\limits_{k}\!\!\!\frac{f_{0}^{\;(k,\;p)}}{(k^{2}-\mu^{2})}=\int\limits_{k}\!\!\!\frac{1}{(k^{2}-\mu^{2})^{2}},
\end{align}
\item Linear divergence
\begin{align}
\int\limits_{k}\!\!\!\frac{f_{1}^{\;(k,\;p)}}{(k^{2}-\mu^{2})}=\int\limits_{k}\!\!\!\frac{2p\cdot k}{(k^{2}-\mu^{2})^{3}},
\end{align}
\item Logarithmic divergence
\begin{align}
\int\limits_{k}\!\!\!\frac{f_{2}^{\;(k,\;p)}}{(k^{2}-\mu^{2})}=\int\limits_{k}\!\!\!\frac{1}{(k^{2}-\mu^{2})^{3}}\!\!\left[\frac{(2p\cdot k)^{2}}{(k^{2}-\mu^{2})}-p^{2}\!\right]\!\!.
\end{align}
\end{enumerate}

The quadratic and linear divergences vanish in the limit $\mu^{2}\rightarrow 
0$. The logarithmic one contains two terms which can be identified with 
$I_{log}^{\mu \nu}(\mu^{2})$ and $I_{log}(\mu^{2})$ respectively. We use 
identity (\ref{identsurface}) to express all divergences in terms of $I_{log}(\mu^{2})$ 
and therefore the amplitude is given by
\begin{align}
\frac{\Xi^{(1)}}{g^{2}}=-\frac{p^{2}}{6}I_{log}(\mu^{2})+2p^{2}\Upsilon_{1}+\mbox{finite},
\label{amplitudep1}
\end{align} 
\noindent
where $\Upsilon_{1}$ is an arbitrary constant steaming from a surface term. The explicit expression for the finite part of (\ref{amplitudep1}) reads
\begin{align}
&\frac{1}{2}\int\limits_{k}\frac{f_{3}^{\;(k,\;p)}+f_{4}^{\;(k,\;p)}+{\bar{f}}^{\;(k,\;p)}}{(k^{2}-\mu^{2})}=\nonumber\\&\frac{1}{2}\int\limits_{k}\!\!\!\frac{1}{(k^{2}-\mu^{2})^{4}}\!\!\left[-4p^2(p\cdot k)+p^{4}
-\!\frac{(p^2-2p \cdot k)^{3}}{(k-p)^2-\mu^2}\!\right]\!\!=\nonumber\\
&=\frac{p^{2}b_{6}}{6}\ln\left(-\frac{p^{2}}{\mu^{2}}\right)-\frac{4p^{2}b_{6}}{9}+O(\mu^2).
\end{align}

Using the scale relation (eq. \ref{scale}) and taking the limit $\mu^{2}\rightarrow 0$ we finally obtain
\begin{align}
\Xi^{(1)}\!=\!-\frac{g^2p^2}{6}\!\!\left[\!I_{log}(\lambda^2)\!-b_{6}\ln\left(\!\!-\frac{p^2}{\lambda^2}\right)\!+\!\frac{8b_{6}}{3}\!-\!12\Upsilon_{1}\!\right]\!\!.
\end{align}

In a similar way we obtain the amplitude of the one-loop correction for the vertex function:

\begin{figure}[h!]
\vspace{-0.21cm}
\begin{center}
\includegraphics{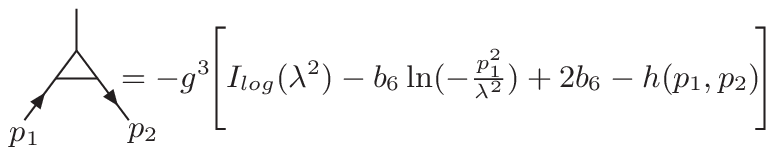}
\end{center}
\vspace{-0.635cm}
\caption{Graph $V^{(1)}$}
\vspace{-0.28cm}
\label{v1}
\end{figure}
\noindent
In the amplitude depicted by $V^{(1)}$, $h(p_{1},p_{2})$ is a function of $p_{1}$ and $p_{2}$ which vanishes if $p_{2}=0$.

Although the previous examples are too simple to display the terms to be subtracted by Bogoliubov's recursion formula (the one-loop graphs do not contain subdivergences), we presented them here to familiarize the reader with the rules of IReg and identity (\ref{identbphz}). We proceed now to the two-loop corrections for the propagator and the vertex functions. The graphs needed in the renormalization of the propagator are:

\begin{figure}[h!]
\begin{center} 
\includegraphics{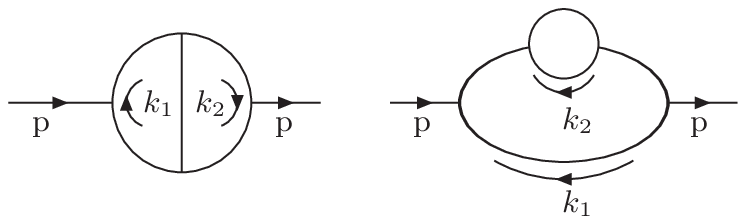}
\end{center}
\vspace{-0.35cm}
\caption{Graphs $P_{A}^{(2)}$ and $P_{B}^{(2)}$ respectively}
\label{p2}
\end{figure}

The amplitude corresponding to $P_{A}^{(2)}$ is given by
\begin{align}
\frac{\Xi_{A}^{(2)}}{ig^4}=\frac{1}{2}\int\limits_{k_{1}k_{2}}\Delta(k_{1})\Delta(k_{1}-p)\Delta(k_{1}-k_{2})\times\nonumber\\\Delta(k_{2})\Delta(k_{2}-p),
\end{align}
\noindent
where we defined
\be
\Delta(k_{i})\equiv\frac{1}{k_{i}^{2}-\mu^2}.
\ee

We begin by applying identity (\ref{identbphz}) in the propagators that depend on the external momenta to obtain
\begin{align}
&\int\limits_{k_{1}k_{2}}\!\!\!\!\Delta(k_{1})\Delta(k_{1}-k_{2})\Delta(k_{2})\times\nonumber\\&\!\!\!\left[\!\sum_{l=0}^{2(n^{(k_{1})}-1)}\!\!\!\!\!f_{l}^{\;(k_{1},\;p)}\!\!+\!{\bar{f}}^{\;(k_{1},\;p)}\!\right]\!\!\!\!\left[\!\sum_{m=0}^{2(n^{(k_{2})}-1)}\!\!\!\!\!f_{m}^{\;(k_{2},\;p)}\!\!+\!{\bar{f}}^{\;(k_{2},\;p)}\!\right]\!\!\!.
\label{intid}
\end{align}

Our next task is to determine the values of $\;n^{(k_{i})}$. They are chosen in order to assure the finitude of the terms that contain ${\bar{f}}^{(k_{i},\;p)}$ as $k_{i}\rightarrow \infty$ in all possible ways. We consider first $\;n^{(k_{1})}$. The terms that contain ${\bar{f}}^{(k_{1},\;p)}$ are
\begin{align}
&\int\limits_{k_{1}k_{2}}\Delta(k_{1})\Delta(k_{1}-k_{2})\Delta(k_{2}){\bar{f}}^{\;(k_{1},\;p)}\times\nonumber\\&\quad\quad\quad\left[\!\sum_{m=0}^{2(n^{(k_{2})}-1)}f_{m}^{\;(k_{2},\;p)}+{\bar{f}}^{\;(k_{2},\;p)}\!\right]=\nonumber\\&\int\limits_{k_{1}k_{2}}\!\!\!\Delta(k_{1})\Delta(k_{1}-k_{2})\Delta(k_{2}){\bar{f}}^{(k_{1},\;p)}\Delta(k_{2}-p).
\end{align} 

We want to guarantee that the above integral is finite as $k_{1}\rightarrow \infty$. There are two cases:

\begin{enumerate}
\item Finitude as $k_{1}\rightarrow \infty$ and $k_{2}$ fixed: $n^{(k_{1})}>0$,
\item Finitude as $k_{1}\rightarrow \infty$ and $k_{2}\rightarrow \infty$: $n^{(k_{1})}>\!2$,
\end{enumerate}
\noindent
which leads us to conclude that $n^{(k_{1})}$ should be at least 3. In a similar fashion, we obtain $n^{(k_{2})}=3$. 

Having found the values of $n^{(k_{i})}$, we proceed to identify the divergent terms contained in (\ref{intid}) as $k_{1}$ and/or $k_{2}$ go to infinity. There are three possibilities. We start with the case $k_{1}\rightarrow \infty$ and $k_{2}$ fixed where the divergence terms are of the type
\begin{align} \int\limits_{k_{1}k_{2}}\!\!\!\!\Delta(k_{1})\Delta(k_{2})\Delta(k_{1}-k_{2})f_{l}^{\;(k_{1},\;p)}\times\nonumber\\\left[\!\sum_{m=0}^{4}f_{m}^{\;(k_{2},\;p)}+{\bar{f}}^{\;(k_{2},\;p)}\!\right].
\end{align}

As $f_{l}^{\;(k_{1},\;p)}$ goes like $k_{1}^{-(l+2)}$, we find by power counting that the divergent terms in this case are given by
\begin{align}
A_{1}^{\Xi}&\equiv\!\!\int\limits_{k_{1}k_{2}}\!\!\Delta(k_{1})\Delta(k_{2})\Delta(k_{1}-k_{2})f_{0}^{\;(k_{1},\;p)}\times\nonumber\\&\quad\quad\quad\quad\quad\quad\left[\!\sum_{m=0}^{4}f_{m}^{\;(k_{2},\;p)}+{\bar{f}}^{\;(k_{2},\;p)}\!\right]=\nonumber\\&=\!\!\int\limits_{k_{1}k_{2}}\!\!\Delta^{2}(k_{1})\Delta(k_{1}-k_{2})\Delta(k_{2})\Delta(k_{2}-p).
\label{1k}
\end{align}

We consider now the case where $k_{2}\rightarrow \infty$ and $k_{1}$ is fixed. Repeating the previous reasoning, we find that the divergent terms are
\begin{align}
A_{2}^{\Xi}&\equiv\!\!\int\limits_{k_{1}k_{2}}\!\!\Delta(k_{1})\Delta(k_{2})\Delta(k_{1}-k_{2})f_{0}^{\;(k_{2},\;p)}\times\nonumber\\&\quad\quad\quad\quad\quad\quad\left[\!\sum_{l=0}^{4}f_{l}^{\;(k_{1},\;p)}+{\bar{f}}^{\;(k_{1},\;p)}\!\right]=\nonumber\\&=\!\!\int\limits_{k_{1}k_{2}}\!\!\Delta^{2}(k_{2})\Delta(k_{1}-k_{2})\Delta(k_{1})\Delta(k_{1}-p).
\label{2k}
\end{align}

Finally we consider $k_{1}\rightarrow \infty$ and $k_{2}\rightarrow \infty$ 
simultaneously. The choice of $n^{(k_{i})}=3$ ($i=1,2$) assures us that the divergent terms must be of the type
\begin{align} \int\limits_{k_{1}k_{2}}\!\!\!\!\Delta(k_{1})\Delta(k_{2})\Delta(k_{1}-k_{2})f_{l}^{\;(k_{1},\;p)}f_{m}^{\;(k_{2},\;p)}.
\end{align}

By power counting, we obtain that $l$ and $m$ are constrained by $l+m\le2$. The cases $l=0$ and $m=0,1,2$ are contained in $A_{1}^{\Xi}$ (eq. \ref{1k}) while the cases $m=0$ and $l=0,1,2$ are contained in $A_{2}^{\Xi}$ (eq. \ref{2k}). We are therefore left with the case $l=m=1$ which reads
\begin{align}
&A_{3}^{\Xi}\equiv\!\!\int\limits_{k_{1}k_{2}}\!\!\Delta(k_{2})\Delta(k_{1}-k_{2})\Delta(k_{1})f_{1}^{\;(k_{1},\;p)}f_{1}^{\;(k_{2},\;p)}=\nonumber\\&\quad\;=\!\!\int\limits_{k_{1}k_{2}}\!\!\Delta^{3}(k_{1})\Delta(k_{1}-k_{2})\Delta^{3}(k_{2})(2p \cdot k_{1})(2p \cdot k_{2}).
\end{align}

Summarizing, the divergent terms are:

\begin{enumerate}
\item Divergences as $k_{1}\rightarrow \infty$ and $k_{2}$ is fixed
\begin{align}
A_{1}^{\Xi}\!=\!\int\limits_{k_{1}k_{2}}\!\!\Delta^{2}(k_{1})\Delta(k_{1}-k_{2})\Delta(k_{2})\Delta(k_{2}-p),
\label{k1}
\end{align}
\item Divergences as $k_{2}\rightarrow \infty$ and $k_{1}$ is fixed
\begin{align}
A_{2}^{\Xi}\!=\!\int\limits_{k_{1}k_{2}}\!\!\Delta^{2}(k_{2})\Delta(k_{1}-k_{2})\Delta(k_{1})\Delta(k_{1}-p),
\label{k2}
\end{align}
\item Divergences as $k_{1}\rightarrow \infty$ and $k_{2}\rightarrow \infty$ simultaneously
\begin{align}
A_{3}^{\Xi}=\!\!\int\limits_{k_{1}k_{2}}\!\!\Delta^{3}(k_{1})\Delta(k_{1}-k_{2})\Delta^{3}(k_{2})(2p \cdot k_{1})(2p \cdot k_{2}).
\label{k1k2}
\end{align}
\end{enumerate}

Therefore, the divergent content of $\Xi_A^{(2)}$ is given by $A_{1}^{\Xi}+A_{2}^{\Xi}+A_{3}^{\Xi}-A_{4}^{\Xi}$. The last term corresponds to the case ($l=m=0$) 
\begin{align}
A_{4}^{\Xi}\equiv&\int\limits_{k_{1}k_{2}}\Delta(k_{2})\Delta(k_{1}-k_{2})\Delta(k_{1})f_{0}^{\;(k_{1},\;p)}f_{0}^{\;(k_{2},\;p)}=\nonumber\\=&\int\limits_{k_{1}k_{2}}\Delta^{2}(k_{2})\Delta(k_{1}-k_{2})\Delta^{2}(k_{1})
\label{k0}
\end{align}
\noindent
and must be subtracted because it is counted twice.

The above classification of the divergent terms in different cases (the term $A_{4}^{\Xi}$ can be thought of as the intersection between the cases $k_{1}\rightarrow \infty$ and $k_2$ fixed, $k_{2}\rightarrow \infty$ and $k_1$ fixed) furnishes us a natural order in which the integrals must be performed and allow us to implement IReg to multi-loop Feynman graphs in a systematic way. More importantly, as a byproduct it also displays the terms to be subtracted by Bogoliubov's recursion formula as we shall verify. Roughly speaking, for each of the cases we have studied characterizing the divergent behaviour in all possible ways that the internal momenta go to infinity we may readily apply the formalism developed in Section \ref{s:overview} to define basic divergent integrals. For this purpose we use identity (\ref{ident}) to each case taking $k_{l}$ to be the internal momentum that we pick to go to infinity. In our example, $k_{1(2)}$ in $A_{1(2)}^{\Xi}$ and $k_{1}$ and $k_{2}$ in $A_{3(4)}^{\Xi}$.

Examining $A_{1}^{\Xi}$ and $A_{2}^{\Xi}$, we notice that both have the same structure and the integral in which we are going to use the rules of IReg is given by 
\begin{align}
\int\limits_{k_{i}}\Delta^{2}(k_{i})\Delta(k_{i}-k_{j}),\quad i,j=1,2\!\!\! \quad\mbox{and}\!\!\!\quad i\neq j
\label{vertex}
\end{align}
 
This is the same amplitude of graph $V_1$ (fig \ref{v1}) if we identify $p_{1}\rightarrow k_{j}$ and set $p_{2}=0$. Thus we can readily write
\begin{align}
A_{i}^{\Xi}&=\bar{A}_{i}^{\Xi}+\alpha_{i}^{\Xi},\quad\quad i,j=1,2 \quad\mbox{and}\quad i\neq j\nonumber\\
\bar{A}_{i}^{\Xi}&\equiv\int\limits_{k_{j}}\Delta(k_{j})\Delta(k_{j}-p)\left[I_{log}(\lambda^2)\right],\nonumber\\\alpha_{i}^{\Xi}&\equiv b_{6}\!\int\limits_{k_{j}}\!\Delta(k_{j})\Delta(k_{j}-p)\ln\left(-\frac{k_{j}^{2}-\mu^2}{e^{2}\lambda^2}\right)^{\!\!-1}\!\!\!\!\!\!.
\label{k1bphz}
\end{align}

We turn to $A_{3}^{\Xi}$. Since the integral in $k_{1}$ is finite, we evaluate it by Feynman parametrization. We insert the result in the integral in $k_{2}$ and use the rules of IReg to obtain
\be
\bar{\alpha}_{3}^{\Xi}\equiv A_{3}^{\Xi}=b_{6}p^{2}\left[\frac{I_{log}(\lambda^2)}{3}+2\Upsilon_{1}\right].
\label{resultk1k2}
\ee
\noindent
Similarly
\begin{align}
A_{4}^{\Xi}&=\int\limits_{k_{2}}\Delta^{2}(k_{2})\left[I_{log}(\lambda^2)-b_{6}\!\ln\left(\!\!-\frac{\!k_{2}^{2}-\!\mu^2}{\lambda^2}\right)\!\!+2b_{6}\!\right]\nonumber\\&=0
\end{align}
\noindent
in the limit $\mu^{2}\rightarrow 0$.

We will show that the terms $\bar{A}_{i}^{\Xi}$ ($i=1,2$) are just the ones which must be subtracted by Bogoliubov's recursion formula. Let us set them aside for the time being and evaluate the rest ($\alpha_{i}^{\Xi}$). Using identity (\ref{identbphz}) in the propagator that depend on the external momentum, we identify the divergent terms. In the limit $\mu^{2}\rightarrow0$, the only one that contributes is
\begin{align}
\bar{\alpha}_{i}^{\Xi}\equiv\int\limits_{k_{j}}\Delta(k_{j})f_{2}^{\;(k_{j},\;p)}\!\!\left[-b_{6}\ln\left(-\frac{k_{j}^{2}-\mu^2}{\lambda^2}\right)+2b_{6}\right]
\end{align}
\noindent
which can be expressed by
\begin{align}
\bar{\alpha}_{i}^{\Xi}=b_{6}p^{2}\Bigg[\frac{I_{log}^{(2)}(\lambda^2)}{3}-\frac{8}{9}I_{log}(\lambda^2)+8\Upsilon_{1}-4\Upsilon_{2}\Bigg].
\end{align}

Hence, the divergent content of $\Xi_A^{(2)}$ plus surface terms is given by
\begin{align}
\frac{\Xi_{A}^{(2)\infty}}{ig^{4}}\equiv\frac{1}{2}\big(\bar{\alpha}_{1}^{\Xi}+\bar{\alpha}_{2}^{\Xi}+\bar{\alpha}_{3}^{\Xi}+\bar{A}_{1}^{\Xi}+\bar{A}_{2}^{\Xi}\big).
\label{divpa2}
\end{align}

The two last terms are just the ones to be subtracted by Bogoliubov's recursion formula. In fact, in order to subtract the subdivergences of this particular graph we must add the following counterterms
\vspace{0.9cm}
\begin{figure}[h!]
\vspace{-0.9cm}
\begin{center} 
\includegraphics{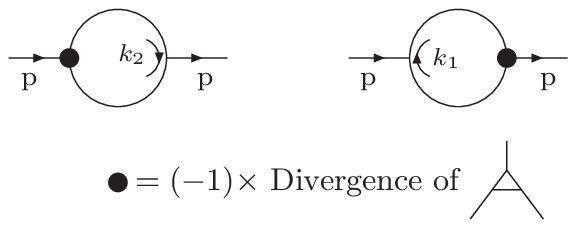}
\end{center}
\vspace{-0.175cm}
\caption{Counterterms for $P_{A}^{(2)}$}
\label{cp2}
\end{figure}

\noindent
whose amplitudes are, respectively
\begin{align}
\frac{ig^{4}}{2}\!\int\limits_{k_{2}}\!\Delta(k_{2})\Delta(k_{2}-p)\left[-I_{log}(\lambda^2)\right]=\frac{ig^{4}}{2}\left(-\bar{A}_{1}^{\Xi}\right),
\end{align}
\begin{align}
\frac{ig^{4}}{2}\!\int\limits_{k_{1}}\!\Delta(k_{1})\Delta(k_{1}-p)\left[-I_{log}(\lambda^2)\right]=\frac{ig^{4}}{2}\left(-\bar{A}_{2}^{\Xi}\right).
\end{align}
Notice that we are adopting a ``MS"\ scheme that, in IReg, corresponds to the subtraction of basic divergent integrals \cite{Sampaio:2002ii}.

Therefore, subtracting the subdivergences yields
\begin{align}
\frac{\bar{\Xi}_A^{(2)}}{ig^4}\equiv\frac{b_{6}p^2}{6}\Bigg[&2I_{log}^{(2)}(\lambda^2)-\frac{13}{3}I_{log}(\lambda^2)+\nonumber\\&\quad+54\Upsilon_{1}-24\Upsilon_{2}+\mbox{finite}\Bigg].
\end{align}

We turn now to the two loop nested graph ($P_{B}^{(2)}$) whose amplitude is
\begin{align}
\frac{\Xi_B^{(2)}}{ig^{4}}=\frac{1}{2}\!\int\limits_{k_{1}k_{2}}\!\!\!\!\Delta^{2}(k_{1})\Delta(k_{1}-p)\Delta(k_{2})\Delta(k_{1}-k_{2}).
\end{align}

We use identity (\ref{identbphz}) in the propagator that depend on the external momentum and find that the choice $n^{(k_{1})}=3$ guarantees the finitude of the terms that contain ${\bar{f}}^{(k_{1},\;p)}$ as $k_{1}\rightarrow \infty$ in all possible ways. We proceed to identify the divergent terms. The case $k_{1}\rightarrow \infty$ and $k_{2}$ fixed does not contain any divergent term while the case $k_{2}\rightarrow \infty$ and $k_{1}$ fixed does. They are given by
\begin{align}
&\!\int\limits_{k_{1}k_{2}}\!\!\!\!\Delta^{2}(k_{1})\Delta(k_{1}-k_{2})\Delta(k_{2})\!\!\left[\sum_{l=0}^{4}f_{l}^{\;(k_{1},\;p)}\!\!+\!{\bar{f}}^{\;(k_{1},\;p)}\!\right]\!\!=\nonumber\\&\int\limits_{k_{1}k_{2}}\!\!\!\!\Delta^{2}(k_{1})\Delta(k_{1}-k_{2})\Delta(k_{2})\Delta(k_{1}-p).
\end{align}

For definiteness call the above integral $B_{1}^{\Xi}$. This is the only one which we have to deal with (the divergent terms from the case $k_{1}\rightarrow \infty$ and $k_{2}\rightarrow \infty$ simultaneously are contained in the above integral). One may notice that it is just the original amplitude of the graph but now we have a natural order to implement IReg. We use its rules in the integral in $k_{2}$ to obtain
\begin{align}
&B_{1}^{\Xi}=\bar{B}_{1}^{\Xi}+\beta_{1}^{\Xi},\nonumber\\
&\bar{B}_{1}^{\Xi}\equiv\int\limits_{k_{1}}\Delta(k_{1})\Delta(k_{1}-p)\left[-\frac{I_{log}}{3}(\lambda^2)\right],\nonumber
\end{align}
\begin{align}
&\beta_{1}^{\Xi}\equiv\int\limits_{k_{1}}\Delta(k_{1})\Delta(k_{1}-p)\Bigg[\frac{b_{6}}{3}\ln\left(-\frac{k_{1}^{2}-\mu^{2}}{\lambda^2}\right)\Bigg]+\nonumber\\&\quad\quad\quad\;\;+\int\limits_{k_{1}}\Delta(k_{1})\Delta(k_{1}-p)\Bigg[4\Upsilon_{1}-\frac{8b_{6}}{9}\Bigg].
\end{align}

We apply the procedure again in $\beta_{1}^{\Xi}$ to find the following divergent terms
\begin{align}
\bar{\beta}_{1}^{\Xi}\equiv\!&\int\limits_{k_{1}}\!\Delta(k_{1})\!\left[\sum_{l=0}^{2}f_{l}^{\;(k_{2},\;p)}\!\right]\!\!\!\left[\frac{b_{6}}{3}\ln\left(\!-\frac{k_{1}^{2}-\mu^{2}}{\lambda^2}\right)\right]\!+\!\nonumber\\&+\int\limits_{k_{1}}\!\Delta(k_{1})\!\left[\sum_{l=0}^{2}f_{l}^{\;(k_{2},\;p)}\!\right]\!\!\left(4\Upsilon_{1}-\frac{8b_{6}}{9}\right)\nonumber\\=\!&
-\frac{p^2}{3}\Bigg\{\frac{b_{6}}{3}I_{log}^{(2)}(\lambda^2)-\frac{10b_{6}}{9}I_{log}(\lambda^2)-4b_{6}\Upsilon_{2}\nonumber\\&\quad\quad\;\;+4\Upsilon_{1}\Bigg[I_{log}(\lambda^2)+\frac{2b_{6}}{3}-3\Upsilon_{1}\Bigg]\!\Bigg\}.
\end{align}

Notice that an arbitrary valued surface term appears multiplied by a divergence expressed by $I_{log}(\lambda^2)$. Should we have adopted CIReg which sets $\Upsilon_{l}=0$ as required by momentum routing invariance (and gauge symmetry) it would not have appeared. We shall keep all surface terms until the end to see whether they play any role in the physics of a less symmetrical theory such as scalar field theories.

Therefore, the divergent content of $\Xi_B^{(2)}$ plus surface terms is given by
\begin{align}
\frac{\Xi_B^{(2)\infty}}{ig^{4}}\equiv\frac{1}{2}\left(\bar{\beta}_{1}^{\Xi}+\bar{B}_{1}^{\Xi}\right).
\end{align}

Again, the last term is just the one which is subtracted by Bogoliubov's recursion formula since the counterterm we must add is
\begin{figure}[ht]
\begin{center} 
\includegraphics{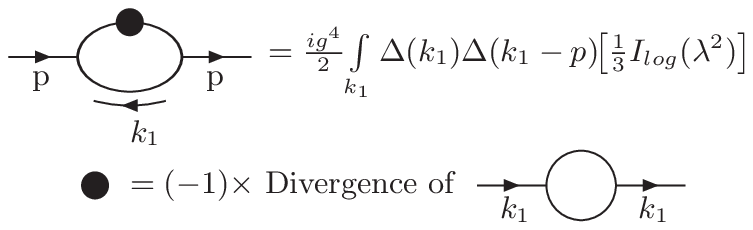}
\end{center}
\caption{Counterterm for $P_{B}^{(2)}$}
\label{c2p2}
\end{figure}

After we add the counterterm we obtain
\begin{align}
&\frac{\bar{\Xi}_B^{(2)}}{ig^{4}}\!\equiv\frac{b_{6}p^2}{6}\Bigg\{-\frac{I_{log}^{(2)}(\lambda^2)}{3}+\frac{10}{9}I_{log}(\lambda^2)+4\Upsilon_{2}\Bigg\}+\nonumber\\&\quad\;\;+\frac{p^2}{6}\Bigg\{\!\!-4\Upsilon_{1}\!\Bigg[I_{log}(\lambda^2)+\frac{2b_{6}}{3}-3\Upsilon_{1}\!\Bigg]\!+\mbox{finite}\!\Bigg\}.
\end{align}

Notice that the term proportional to $4\Upsilon_{1}I_{log}(\lambda^2)$ is not subtracted as a subdivergence since we are adopting a ``MS" scheme in IReg. In the next section we discuss what would happen if we had included the surface term in the counterterm.

At this point, we can write down the renormalization of the propagator at two loop order
\begin{align}
\bar{\Xi}^{(2)}&\equiv\bar{\Xi}_A^{(2)}+\bar{\Xi}_B^{(2)}\nonumber\\&=ig^4\frac{p^2}{6}\Bigg\{\!\frac{5b_{6}}{3}I_{log}^{(2)}(\lambda^2)\!-\!\frac{29b_{6}}{9}I_{log}(\lambda^2)\!-20\Upsilon_{2}+\nonumber\\&\quad-2\Upsilon_{1}\left[2I_{log}(\lambda^2)-\frac{77b_{6}}{3}+6\Upsilon_{1}\right]+\mbox{finite}\Bigg\}.
\label{prop}
\end{align}

We now turn to the renormalization of the vertex. The graphs we need to evaluate are:
\vspace{2cm}
\begin{figure}[ht]
\begin{center} 
\includegraphics{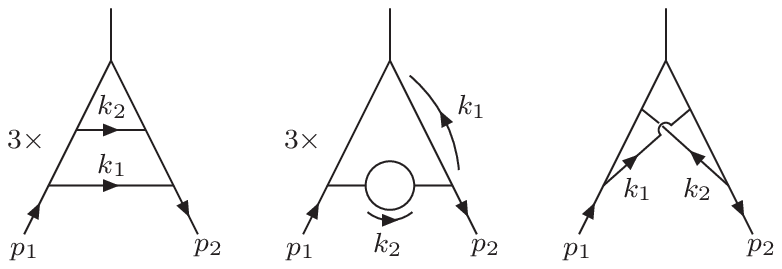}
\vspace{-0.8cm}
\end{center}
\caption{Graphs $V_{A}^{(2)}$, $V_{B}^{(2)}$ and $V_{C}^{(2)}$ respectively}
\end{figure}

The amplitude depicted by $V_{A}^{(2)}$ is given by
\begin{align}
\frac{\Lambda_{A}^{(2)}}{-ig^{5}}\!\equiv\!\!\!\!\int\limits_{k_{1}k_{2}}\!\!\!\!\Delta(k_{1})\Delta(k_{2}\!-k_{1})\!\prod_{i=1}^{2}\!\Delta(k_{i}\!-p_{1})\Delta(k_{i}-p_{2}).
\end{align}

We apply identity (\ref{identbphz}) in the propagators that depend on the external momenta and make the choice $n^{(k_{1})}_{1}=n^{(k_{1})}_{2}=n^{(k_{2})}_{1}=n^{(k_{2})}_{2}=1$ in the usual way to obtain
\begin{align}
\int\limits_{k_{1}k_{2}}&\Delta(k_{1})\Delta(k_{2}-k_{1})\left[f_{0}^{\;(k_{1},\;p_{1})}+{\bar{f}}^{\;(k_{1},\;p_{1})}\right]\times\nonumber\\&\left[f_{0}^{\;(k_{1},\;p_{2})}+{\bar{f}}^{\;(k_{1},\;p_{2})}\right]\!\!\left[f_{0}^{\;(k_{2},\;p_{1})}+{\bar{f}}^{\;(k_{2},\;p_{1})}\right]\times\nonumber\\&\left[f_{0}^{\;(k_{2},\;p_{2})}+{\bar{f}}^{\;(k_{2},\;p_{2})}\right].
\end{align}

The divergent terms comes only from the case $k_{2}\rightarrow \infty$ and $k_{1}$ fixed and they are given by
\begin{align}
A_{1}^{\Lambda}\!&\equiv\!\!\int\limits_{k_{1}k_{2}}\!\!\Delta(k_{1})\Delta(k_{2}-k_{1})f_{0}^{\;(k_{2},\;p_{1})}f_{0}^{\;(k_{2},\;p_{2})}\times\nonumber\\&\quad\quad\left[f_{0}^{\;(k_{1},\;p_{1})}+{\bar{f}}^{\;(k_{1},\;p_{1})}\right]\!\!\left[f_{0}^{\;(k_{1},\;p_{2})}+{\bar{f}}^{\;(k_{1},\;p_{2})}\right]\nonumber\\&=\!\!\!\!\int\limits_{k_{1}k_{2}}\!\!\!\!\Delta(k_{1})\Delta(k_{2}\!-\!k_{1})\Delta^2(k_{2})\Delta(k_{1}\!-\!p_{1})\Delta(k_{1}\!-\!p_{2})\nonumber\\
&=\bar{A}_{1}^{\Lambda}+\alpha_{1}^{\Lambda},\nonumber\\
\bar{A}_{1}^{\Lambda}\!&\equiv\int\limits_{k_{1}}\Delta(k_{1})\Delta(k_{1}-p_{1})\Delta(k_{1}-p_{2})\left[I_{log}(\lambda^2)\right],\nonumber\\
\alpha_{1}^{\Lambda}\!&\equiv\!\! \int\limits_{k_{1}}\!\!\Delta(k_{1})\Delta(k_{1}\!-\!p_{1}\!)\Delta(k_{1}\!-\!p_{2})\ln\!\left(\!-\frac{k_{1}^{2}-\mu^2}{e^{2}\lambda^2}\right)^{\!\!\!-b_{6}}\!\!\!\!\!\!.
\label{vertexk1}
\end{align}

We apply the procedure again in $\alpha_{1}^{\Lambda}$ to yield the divergent term below
\begin{align}
\bar{\alpha}_{1}^{\Lambda}\!&\equiv b_{6}\!\int\limits_{k_{1}}\!\!\!\Delta(k_{1})f_{0}^{\;(k_{1},\;p_{1})}\!f_{0}^{\;(k_{1},\;p_{2})}\!\!
\left[2\!-\!\ln\!\left(\!-\frac{k_{1}^{2}-\!\mu^2}{\lambda^2}\right)\!\right]\nonumber\\&=2b_{6}I_{log}(\lambda^2)-b_{6}I_{log}^{(2)}(\lambda^2).
\end{align}

Hence, the divergent content of $\Lambda_{A}^{(2)}$ is given by
\begin{align}
\Lambda_{A}^{(2)\infty}\equiv-ig^{5}\left[\bar{\alpha}_{1}^{\Lambda}+\bar{A}_{1}^{\Lambda}\right]
\end{align}
\noindent
where the last term is cancelled by the counterterm
\begin{figure}[ht]
\vspace{-0.6cm}
\begin{center} 
\includegraphics{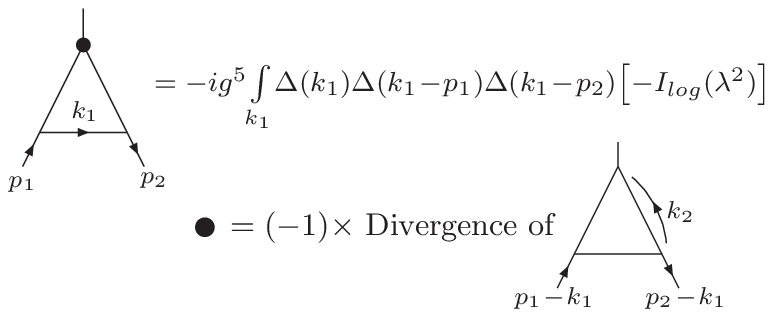}
\vspace{-0.8cm}
\end{center}
\caption{Counterterm for $V_{A}^{(2)}$}
\end{figure}

Therefore, after the subtraction of the subdivergence we have
\begin{align}
\frac{\bar{\Lambda}_{A}^{(2)}}{-ig^{5}}\equiv b_{6}\left[-I_{log}^{(2)}(\lambda^2)+2I_{log}(\lambda^2)+\mbox{finite}\right].
\end{align}

We turn to graph $V_{B}^{(2)}$ whose amplitude is 
\begin{align}
\frac{\Lambda_{B}^{(2)}}{-ig^{5}}\equiv\frac{1}{2}\int\limits_{k_{1}k_{2}}\Delta^{2}(k_{1})&\Delta(k_{1}-p_{1})\Delta(k_{1}-p_{2})\times\nonumber\\&\times\Delta(k_{2})\Delta(k_{2}-k_{1}).
\end{align}

We use identity (\ref{identbphz}) and choose $n^{(k_{1})}_{1}\!\!=n^{(k_{1})}_{2}\!\!=1$. The divergent terms are all contained in the case $k_{2}\rightarrow \infty$ and $k_{1}$ fixed and they are given by
\begin{align}
B_{1}^{\Lambda}\!\equiv&\!\!\!\int\limits_{k_{1}k_{2}}\!\!\!\!\Delta^{2}(k_{1})\Delta(k_{2})\Delta(k_{2}-k_{1})\times\nonumber\\&\quad\!\!\left[f_{0}^{\;(k_{1},\;p_{1})}+{\bar{f}}^{\;(k_{1},\;p_{1})}\right]\!\!\left[f_{0}^{\;(k_{1},\;p_{2})}+{\bar{f}}^{\;(k_{1},\;p_{2})}\right]\nonumber\\=&\!\!\!\!\int\limits_{k_{1}k_{2}}\!\!\!\!\!\Delta^{2}(k_{1})\Delta(k_{2})\Delta(k_{2}\!-\!k_{1})\Delta(k_{1}\!-\!p_{1})\Delta(k_{1}\!-\!p_{2})\nonumber\\
=&\quad\!\!\!\!\bar{B}_{1}^{\Lambda}+\beta_{1}^{\Lambda},\nonumber\\
\bar{B}_{1}^{\Lambda}\equiv&\int\limits_{k_{1}}\Delta(k_{1})\Delta(k_{1}-p_{1})\Delta(k_{1}-p_{2})\left[-\frac{I_{log}}{3}(\lambda^2)\right]\!\!,\nonumber\\\beta_{1}^{\Lambda}\equiv&\int\limits_{k_{1}}\Delta(k_{1})\Delta(k_{1}-p_{1})\Delta(k_{1}-p_{2})\times\nonumber\\&\quad\quad\!\!\!\!\left[4\Upsilon_{1}+\frac{b_{6}}{3}\ln\left(-\frac{k_{1}^{2}-\mu^{2}}{\lambda^2}\right)-\frac{8b_{6}}{9}\right]\!\!.
\end{align}

Repeating the procedure in $\beta_{1}^{\Lambda}$ yields
\begin{align}
\bar{\beta}_{1}^{\Lambda}\equiv&\int\limits_{k_{1}}\Delta(k_{1})\prod\limits_{i=1}^{2}f_{0}^{\;(k_{1},\;p_{i})}\left[\frac{b_{6}}{3}\ln\left(-\frac{k_{1}^{2}-\mu^{2}}{\lambda^2}\right)\right]+\nonumber\\&\quad\quad+\int\limits_{k_{1}}\Delta(k_{1})\prod\limits_{i=1}^{2}f_{0}^{\;(k_{1},\;p_{i})}\left[4\Upsilon_{1}-\frac{8b_{6}}{9}\right]\nonumber\\
=&\frac{b_{6}}{3}I_{log}^{(2)}(\lambda^{2})\!+4\Upsilon_{1}I_{log}(\lambda^{2})-\!\frac{8b_{6}}{9}I_{log}(\lambda^{2})
\end{align}
\noindent
and, therefore, the divergent content of $\Lambda_{B}^{(2)}$ plus surface terms is given by
\begin{align}
\frac{\Lambda_{B}^{(2)\infty}}{-ig^{5}}\equiv\frac{1}{2}(\bar{\beta}_{1}^{\Lambda}+\bar{B}_{1}^{\Lambda}).
\end{align}

The last term is subtracted by Bogoliubov's recursion formula since the counterterm for 
this graph is
\begin{figure}[h!]
\vspace{-0.4cm}
\begin{center} 
\includegraphics{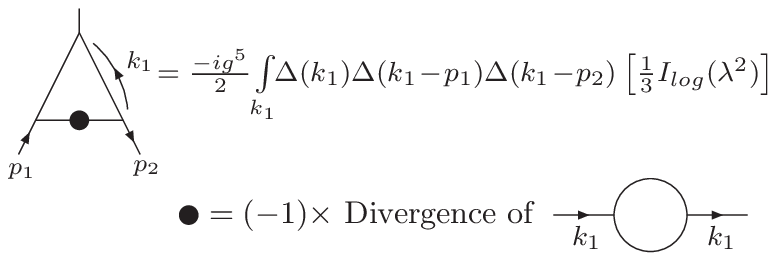}
\end{center}
\vspace{-0.4cm}
\caption{Counterterm for $V_{B}^{(2)}$}
\end{figure}

Subtracting the subdivergence we have
\begin{align}
\frac{\bar{\Lambda}_{B}^{(2)}}{-ig^{5}}\equiv\!\frac{b_{6}}{6}I_{log}^{(2)}(\lambda^2)-\!\left(\!\frac{4b_{6}}{9}-2\Upsilon_{1}\!\right)\!I_{log}(\lambda^2)+\mbox{finite}.
\end{align} 

Finally, we are going to evaluate the graph $V_{C}^{(2)}$. Calling its amplitude $\Lambda_{C}^{(2)}$ we obtain
\begin{align}
\frac{\Lambda_{C}^{(2)}}{-ig^{5}}\equiv\frac{1}{2}\int\limits_{k_{1}k_{2}}\!\!\!\!\Delta(k_{1})\Delta(k_{2})\Delta(k_{1}-p_{1})\Delta(k_{2}-p_{2})\times\nonumber\\\Delta(k_{1}+k_{2}-p_{1})\Delta(k_{1}+k_{2}-p_{2}).
\end{align}

As usual, we choose $n^{(k_{1})}_{1}=n^{(k_{2})}_{2}=n^{(k_{1}+k_{2})}_{1}=n^{(k_{1}+k_{2})}_{2}=1$ and find that the only divergent term comes from the case $k_{1}\rightarrow \infty$ and $k_{2}\rightarrow \infty$ simultaneously and it is given by
\begin{align}
C_{1}^{\Lambda}&\equiv\int\limits_{k_{1}k_{2}}f_{0}^{\;(k_{1},\;p_{1})}f_{0}^{\;(k_{2},\;p_{2})}f_{0}^{\;(k_{1}+k_{2},\;p_{1})}f_{0}^{\;(k_{1}+k_{2},\;p_{2})}\nonumber\\&=\int\limits_{k_{1}k_{2}}\Delta^{2}(k_{1})\Delta^{2}(k_{2})\Delta^{2}(k_{1}+k_{2}).
\end{align}

After the integration over $k_{2}$ and the use of the rules of IReg we have
\begin{align}
\bar{\Lambda}_{C}^{(2)}\equiv\Lambda_{C}^{(2)}=-ig^{5}b_{6}\left[I_{log}(\lambda^2)+\mbox{finite}\right].
\end{align}

Collecting all the results we find that the renormalization of the vertex at two loop order is given by
\begin{align}
\bar{\Lambda}^{(2)}\equiv&\bar{\Lambda}^{(2)}_{A}+\bar{\Lambda}^{(2)}_{B}+\bar{\Lambda}^{(2)}_{C}\nonumber\\=&ig^{5}\!\!\left[\frac{5b_{6}}{2}I_{log}^{(2)}(\lambda^2)\!-\!\left(\frac{17b_{6}}{3}+6\Upsilon_{1}\right)I_{log}(\lambda^2)\right]\!+\nonumber\\&+ig^{5}\times\left(\mbox{finite}\right).
\label{vert}
\end{align}

Although our procedure found success in all the examples already presented, one may wonder how general it is. To answer this question, we apply it to diagrams with more than two-loops.

\subsection{General Algorithm}

In order to see the generality of our method we consider the graph below
\begin{figure}[h!]
\begin{center} 
\includegraphics{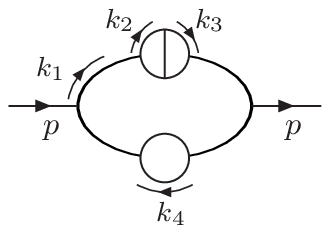}
\end{center}
\vspace{-0.7cm}
\end{figure}

\noindent
whose amplitude is given by
\begin{align}
\frac{\Xi^{(4)}_{A}}{-ig^{8}}\equiv\frac{1}{4}\int\limits_{k_{1}\cdots k_{4}}\!\!\!\!\!
\Delta^{2}&(k_{1})\Delta^{2}(k_{1}\!-\!p)\Delta(k_{2})\Delta(k_{2}\!-\!k_{1})\times\nonumber\\&\Delta(k_{2}-k_{3})\Delta(k_{3})\Delta(k_{3}-k_{1})\times\nonumber\\&\Delta(k_{4})\Delta(k_{4}-k_{1}+p).
\label{termcom}
\end{align}

As usual, we use identity (\ref{identbphz}) in the propagators which depend on the external momenta and choose $n^{(k_{1})}=n^{(k_{4}-k_{1})}=3$. Next, we must identify the divergent terms as the internal momenta go to infinity in all possible ways. For this particular graph, we are lead to an ambiguity because the cases $k_{2(3)}\rightarrow \infty$ and $k_{1},k_{3(2)},k_{4}$ fixed; $k_{2},k_{3}\rightarrow \infty$ and $k_{1},k_{4}$ fixed contain the same divergent terms which read
\begin{align}
\int\limits_{k_{1}\cdots k_{4}}
&\Delta^{2}(k_{1})\left[\sum_{l=0}^{4}f_{l}^{\;(k_{1},\;p)}+{\bar{f}}^{\;(k_{1},\;p)}\!\right]^{2}\Delta(k_{2})\times\nonumber\\&\Delta(k_{2}\!-\!k_{1})\Delta(k_{2}-k_{3})\Delta(k_{3})\Delta(k_{3}-k_{1})\times\nonumber\\&\Delta(k_{4})\left[\sum_{r=0}^{4}f_{r}^{\;(k_{4}-k_{1},\;p)}+{\bar{f}}^{\;(k_{4}-k_{1},\;p)}\!\right].
\end{align}

Therefore, it is not clear if we must sum all of them (and subtract the terms counted twice) or if we must consider just one of the cases. However, it is not the main problem. Supposing we agree what case(s) we must consider, our next task would be to apply the rules of IReg (as presented in Section \ref{s:overview}) according to the previous classification. In any of the cases, the integrals in $k_{2(3)}$ are performed and we are unable to identify the counterterm
\begin{figure}[h!]
\begin{center}
\includegraphics{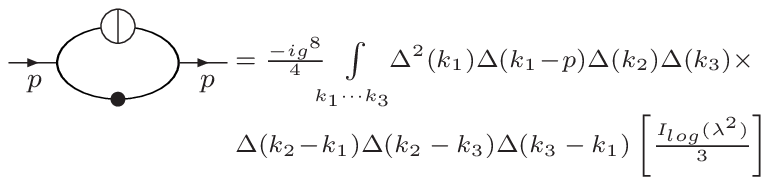}
\end{center}
\vspace{-0.55cm}
\end{figure}

Hence, we need to develop a more general procedure which will contain steps \ref{pro}-\ref{save} as a subcase. We present this new procedure (which can be casted as an algorithm) and, afterwards, explain it step-by-step through the previous example. The algorithm reads
\vspace{-0.3cm}
\begin{figure}[ht]
\begin{center}
\includegraphics{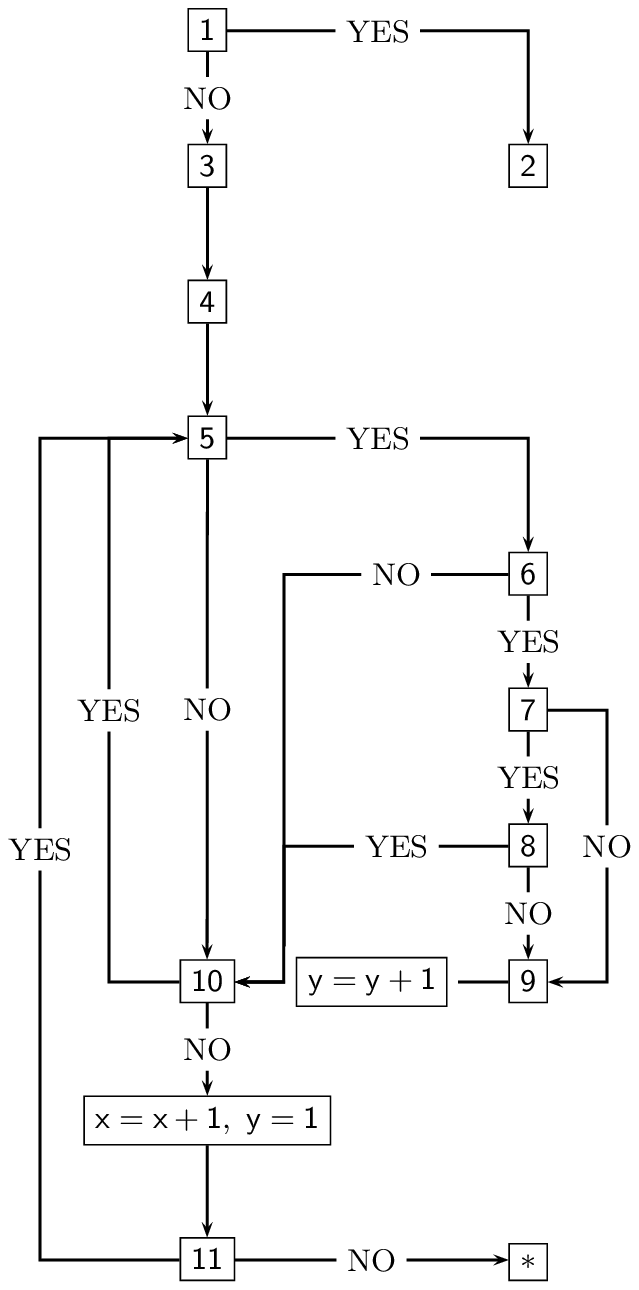}
\end{center}
\end{figure}

\vspace{2.0cm}
\begin{enumerate}
\item Let ${\cal{S}}_{e}$ be the set of subgraphs $G_{i}$ that share one external leg with the whole graph $G$. Let $\bar{{\cal{S}}}_{e}$ be the complementary set. Is $\bar{{\cal{S}}}_{e}$ an empty set?
\label{sub}
\item Apply the procedure stated before (steps \ref{pro}$\rightarrow$\ref{save}) to the amplitude.
\label{AF}
\item Identify the propagators which depend on the external momenta and do not belong to a subgraph contained in $\bar{{\cal{S}}}_{e}$. Use identity (\ref{identbphz}) in such propagators and find out the values of $n_{j}^{(k_{i})}$ in the usual way.
\label{identific}
\item Set $x=0$ and $y=1$. Group all internal momenta in the set $\bar{{\cal{K}}}_{x=0}^{(y=1)}$ and define its complement (which is an empty set) by ${\cal{K}}_{x=0}^{(y=1)}$.
\label{K0}
\item Is there a subgraph $G_{i}$ whose internal momenta are precisely the ones grouped in $\bar{{\cal{K}}}_{x}^{(y)}$?
\label{subyes?}
\item Are there divergent terms as all internal momenta contained in $\bar{{\cal{K}}}_{x}^{(y)}$ go to infinity and all elements of ${\cal{K}}_{x}^{(y)}$ are kept fixed? 
\label{divv}
\item Have all the divergent terms identified in step \ref{divv} been stored?
\label{divok?}
\item Is $\bar{{\cal{K}}}_{x}^{(y)}$ a subset of a already stored $\bar{{\cal{K}}}_{i}^{(j)}$ ($i=0,1,\cdots , x-1$)?
\label{Ksub?}
\item Store the terms identified in step \ref{divv} as well as $\bar{{\cal{K}}}_{x}^{(y)}$ (if $\bar{{\cal{K}}}_{x}^{(y)}$ is a subset of a already stored $\bar{{\cal{K}}}_{i}^{(j)}$, erase the result corresponding to the latter).
\label{store}
\item Choose a new combination of $x$ numbers of internal momenta and group them in the set ${\cal{K}}_{x}^{(y)}$. Is it possible?
\label{choose1}
\item Choose $x$ numbers of internal momenta and group them in the set ${\cal{K}}_{x}^{(y)}$. Is it possible?
\label{choose2}
\end{enumerate}

When the algorithm gets to asterisk ($*$) , we write down all the results stored. Each one is classified according to a particular set of internal momenta that go to infinity. If this set has two or more elements, we identify the subgraph that contains them and run the algorithm on it. On the other hand, if the set has only one element we use the rules of IReg (as presented in Section \ref{s:overview}). At this point, we set aside the terms that contain $I^{(l)}_{log}(\lambda^2)$ and apply the algorithm on the others. Eventually, we obtain two kind of terms: the ones in which $I^{(l)}_{log}(\lambda^2)$ multiplies an integral (that correspond to the terms cancelled by Bogoliubov's recursion formula) and the ones in which $I^{(l)}_{log}(\lambda^2)$ multiplies only constants and/or polynomials in the external momenta (that correspond to the typical divergence of the graph).

As stated, we are going to clarify the method with the previous example. In the following, we present the result we obtain after each one of the steps.

\

$\bullet$ Step \pw{\ref{sub}} : answer is NO
\begin{figure}[h!]
\begin{flushleft}
\includegraphics{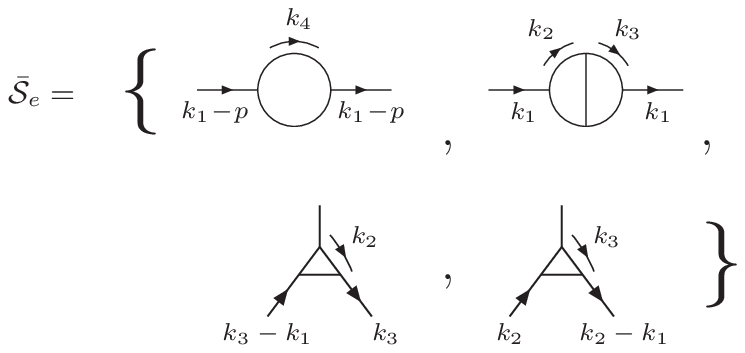}
\end{flushleft}
\end{figure}
\begin{figure}[h!]
\vspace{-0.5cm}
\begin{flushleft}
\includegraphics{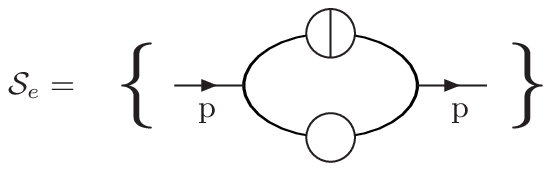}
\end{flushleft}
\end{figure}

$\bullet$ Step \pw{\ref{identific}}
\begin{align}
n^{(k_{1})}=3,\quad\Delta{(k_{1}-p)}=\left[\sum_{l=0}^{4}f_{l}^{\;(k_{1},\;p)}+{\bar{f}}^{\;(k_{1},\;p)}\!\right].\nonumber
\end{align}

$\bullet$ Step \pw{\ref{K0}}
\begin{equation}
\bar{{\cal{K}}}_{0}^{(1)}=\{k_{1}\cdots k_{4}\},\quad {\cal{K}}_{0}^{(1)}=\emptyset\nonumber
\end{equation}

$\bullet$ Step \pw{\ref{subyes?}} : answer is YES

\

The internal momenta $k_{1}\cdots k_{4}$ (which are contained in $\bar{{\cal{K}}}_{x=0}^{(y=1)}$) belong to a subgraph of the set ${\cal{S}}_{e}$.

\

$\bullet$ Step \pw{\ref{divv}} : answer is YES

\

The divergent terms are 
\begin{align}
\int\limits_{k_{1}\cdots k_{4}}
&\Delta^{2}(k_{1})\left[\sum_{l,\;l'=\;0}^{l+l'\leq2}f_{l}^{\;(k_{1},\;p)}f_{l'}^{\;(k_{1},\;p)}\right]\Delta(k_{2})\times\nonumber\\&\Delta(k_{2}\!-\!k_{1})\Delta(k_{2}-k_{3})\Delta(k_{3})\Delta(k_{3}-k_{1})\times\nonumber\\&\Delta(k_{4})\Delta(k_{4}-k_{1}+p).
\label{termident}
\end{align}

$\bullet$ Step \pw{\ref{divok?}} : answer is NO

\

$\bullet$ Step \pw{\ref{store}}

\

We store the divergent terms given by (\ref{termident}) as well as $\bar{{\cal{K}}}_{x=0}^{(y=1)}=\{k_{1}\cdots k_{4}\}$.

\

$\bullet$ Step \pw{\ref{choose1}} : answer is NO

\

$\bullet$ Step \pw{\ref{choose2}} : answer is YES

\

We choose $k_{1}$ and obtain
\begin{align}
\bar{{\cal{K}}}_{x=1}^{(y=1)}=\{k_{2}\cdots k_{4}\},\quad {\cal{K}}_{x=1}^{(y=1)}=\{k_{1}\}\nonumber
\end{align}

$\bullet$ Step \pw{\ref{subyes?}} : answer is NO

\

None subgraph contains $\{k_{2}\cdots k_{4}\}$ as its internal momenta.

\

$\bullet$ Step \pw{\ref{choose1}} : answer is YES

\

We choose $k_{2}$ and obtain
\begin{align}
\bar{{\cal{K}}}_{x=1}^{(y=1)}=\{k_{1},k_{3},k_{4}\},\quad {\cal{K}}_{x=1}^{(y=1)}=\{k_{2}\}\nonumber
\end{align}

$\bullet$ Step \pw{\ref{subyes?}} : answer is NO

\

None subgraph contains $\{k_{1},k_{3},k_{4}\}$ as its internal momenta..

\

$\bullet$ Step \pw{\ref{choose1}} : answer is YES

\

We choose $k_{3}$ and obtain
\begin{align}
\bar{{\cal{K}}}_{x=1}^{(y=1)}=\{k_{1},k_{2},k_{4}\},\quad {\cal{K}}_{x=1}^{(y=1)}=\{k_{3}\}\nonumber
\end{align}

$\bullet$ Step \pw{\ref{subyes?}} : answer is NO

\

None subgraph contains $\{k_{1},k_{2},k_{4}\}$ as its internal momenta.

\

$\bullet$ Step \pw{\ref{choose1}} : answer is YES

\

We choose $k_{4}$ and obtain
\begin{align}
\bar{{\cal{K}}}_{x=1}^{(y=1)}=\{k_{1}\cdots k_{3}\},\quad {\cal{K}}_{x=1}^{(y=1)}=\{k_{4}\}\nonumber
\end{align}

$\bullet$ Step \pw{\ref{subyes?}} : answer is NO

\

None subgraph contains $\{k_{1}\cdots k_{3}\}$ as its internal momenta.

\

$\bullet$ Step \pw{\ref{choose1}} : answer is NO

\

$\bullet$ Step \pw{\ref{choose2}} : answer is YES

\

We choose $k_{1},k_{2}$ and obtain
\begin{align}
\bar{{\cal{K}}}_{x=2}^{(y=1)}=\{k_{3},k_{4}\},\quad {\cal{K}}_{x=2}^{(y=1)}=\{k_{1},k_{2}\}\nonumber
\end{align}

$\bullet$ Step \pw{\ref{subyes?}} : answer is NO

\

None subgraph contains $\{k_{3},k_{4}\}$ as its internal momenta.

\

$\bullet$ Step \pw{\ref{choose1}} : answer is YES

\

We choose $k_{1},k_{3}$ and obtain
\begin{align}
\bar{{\cal{K}}}_{x=2}^{(y=1)}=\{k_{2},k_{4}\},\quad {\cal{K}}_{x=2}^{(y=1)}=\{k_{1},k_{3}\}\nonumber
\end{align}

\

$\bullet$ Step \pw{\ref{subyes?}} : answer is NO

\

None subgraph contains $\{k_{2},k_{4}\}$ as its internal momenta.

\

$\bullet$ Step \pw{\ref{choose1}} : answer is YES

\

We choose $k_{1},k_{4}$ and obtain
\begin{align}
\bar{{\cal{K}}}_{x=2}^{(y=1)}=\{k_{2},k_{3}\},\quad {\cal{K}}_{x=2}^{(y=1)}=\{k_{1},k_{4}\}\nonumber
\end{align}

$\bullet$ Step \pw{\ref{subyes?}} : answer is YES

\

The second subgraph of $\bar{{\cal{S}}}_{e}$ has $\{k_{2},k_{3}\}$ as its internal momenta.

\

$\bullet$ Step \pw{\ref{divv}} : answer is YES

\

The divergent terms are 
\begin{align}
\int\limits_{k_{1}\cdots k_{4}}
&\Delta^{2}(k_{1})\left[\sum_{l=0}^{4}f_{l}^{\;(k_{1},\;p)}+{\bar{f}}^{\;(k_{1},\;p)}\!\right]^{2}\Delta(k_{2})\times\nonumber\\&\Delta(k_{2}\!-\!k_{1})\Delta(k_{2}-k_{3})\Delta(k_{3})\Delta(k_{3}-k_{1})\times\nonumber\\&\Delta(k_{4})\Delta(k_{4}-k_{1}+p).
\label{diver}
\end{align}

$\bullet$ Step \pw{\ref{divok?}} : answer is NO

\

One may notice that (\ref{diver}) possesses more terms than (\ref{termident}).

\

$\bullet$ Step \pw{\ref{store}}

\

Since $\bar{{\cal{K}}}_{x=2}^{(y=1)}$ is a subset of $\bar{{\cal{K}}}_{x=0}^{(y=1)}$, we erase the result corresponding to the latter. Therefore, we store only the divergent terms given by (\ref{diver}) as well as $\bar{{\cal{K}}}_{x=2}^{(y=1)}=\{k_{2},k_{3}\}$.

\

$\bullet$ Step \pw{\ref{choose1}} : answer is YES

\

We choose $k_{2},k_{3}$ and obtain
\begin{align}
\bar{{\cal{K}}}_{x=2}^{(y=2)}=\{k_{1},k_{4}\},\quad {\cal{K}}_{x=2}^{(y=1)}=\{k_{2},k_{3}\}\nonumber
\end{align}

\

$\bullet$ Step \pw{\ref{subyes?}} : answer is NO

\

None subgraph contains $\{k_{1},k_{4}\}$ as its internal momenta.

\

$\bullet$ Step \pw{\ref{choose1}} : answer is YES

\

We choose $k_{2},k_{4}$ and obtain
\begin{align}
\bar{{\cal{K}}}_{x=2}^{(y=2)}=\{k_{1},k_{3}\},\quad {\cal{K}}_{x=2}^{(y=1)}=\{k_{2},k_{4}\}\nonumber
\end{align}

\

$\bullet$ Step \pw{\ref{subyes?}} : answer is NO

\

None subgraph contains $\{k_{1},k_{3}\}$ as its internal momenta.

\

$\bullet$ Step \pw{\ref{choose1}} : answer is YES

\

We choose $k_{3},k_{4}$ and obtain
\begin{align}
\bar{{\cal{K}}}_{x=2}^{(y=2)}=\{k_{1},k_{2}\},\quad {\cal{K}}_{x=2}^{(y=2)}=\{k_{3},k_{4}\}\nonumber
\end{align}

\

$\bullet$ Step \pw{\ref{subyes?}} : answer is NO

\

None subgraph contains $\{k_{1},k_{2}\}$ as its internal momenta.

\

$\bullet$ Step \pw{\ref{choose1}} : answer is NO

\

$\bullet$ Step \pw{\ref{choose2}} : answer is YES

\

We choose $k_{1},k_{2},k_{3}$ and obtain
\begin{align}
\bar{{\cal{K}}}_{x=3}^{(y=1)}=\{k_{4}\},\quad {\cal{K}}_{x=3}^{(y=1)}=\{k_{1},k_{2},k_{3}\}\nonumber
\end{align}

$\bullet$ Step \pw{\ref{subyes?}} : answer is YES

\

The first subgraph of $\bar{{\cal{S}}}_{e}$ has $\{k_{4}\}$ as its internal momentum.

\

$\bullet$ Step \pw{\ref{divv}} : answer is YES

\

The divergent terms are 
\begin{align}
\int\limits_{k_{1}\cdots k_{4}}
&\Delta^{2}(k_{1})\left[\sum_{l=0}^{4}f_{l}^{\;(k_{1},\;p)}+{\bar{f}}^{\;(k_{1},\;p)}\!\right]^{2}\Delta(k_{2})\times\nonumber\\&\Delta(k_{2}\!-\!k_{1})\Delta(k_{2}-k_{3})\Delta(k_{3})\Delta(k_{3}-k_{1})\times\nonumber\\&\Delta(k_{4})\Delta(k_{4}-k_{1}+p).
\label{diver1}
\end{align}

$\bullet$ Step \pw{\ref{divok?}} : answer is YES

\

One may notice that (\ref{diver1}) contains the same terms of (\ref{diver}).

\

$\bullet$ Step \pw{\ref{Ksub?}} : answer is NO

\

Since we erased the result corresponding to $\bar{{\cal{K}}}_{x=0}^{(y=1)}$, we notice that $\bar{{\cal{K}}}_{x=3}^{(y=1)}$ is not a subset of a previously stored $\bar{{\cal{K}}}_{x=i}^{(y=j)}$.

\

$\bullet$ Step \pw{\ref{store}}

\

We store the divergent terms given by (\ref{diver1}) as well as $\bar{{\cal{K}}}_{x=3}^{(y=1)}=\{k_{4}\}$.

\

$\bullet$ Step \pw{\ref{choose1}} : answer is YES

\

We choose $k_{1},k_{2},k_{4}$ and obtain
\begin{align}
\bar{{\cal{K}}}_{x=3}^{(y=2)}=\{k_{3}\},\quad {\cal{K}}_{x=3}^{(y=2)}=\{k_{1},k_{2},k_{4}\}\nonumber
\end{align}

\

$\bullet$ Step \pw{\ref{subyes?}} : answer is YES

\

The fourth subgraph of $\bar{{\cal{S}}}_{e}$ has $\{k_{3}\}$ as its internal momentum.

\

$\bullet$ Step \pw{\ref{divv}} : answer is YES

\

The divergent terms are 
\begin{align}
\int\limits_{k_{1}\cdots k_{4}}
&\Delta^{2}(k_{1})\left[\sum_{l=0}^{4}f_{l}^{\;(k_{1},\;p)}+{\bar{f}}^{\;(k_{1},\;p)}\!\right]^{2}\Delta(k_{2})\times\nonumber\\&\Delta(k_{2}\!-\!k_{1})\Delta(k_{2}-k_{3})\Delta(k_{3})\Delta(k_{3}-k_{1})\times\nonumber\\&\Delta(k_{4})\Delta(k_{4}-k_{1}+p).
\label{diver2}
\end{align}

$\bullet$ Step \pw{\ref{divok?}} : answer is YES

\

One may notice that (\ref{diver2}) contains the same terms of (\ref{diver}) and (\ref{diver1}).

\

$\bullet$ Step \pw{\ref{Ksub?}} : answer is YES

\

One may notice that $\bar{{\cal{K}}}_{x=3}^{(y=2)}$ is a subset of $\bar{{\cal{K}}}_{x=2}^{(y=1)}$.

\

$\bullet$ Step \pw{\ref{choose1}} : answer is YES

\

We choose $k_{1},k_{3},k_{4}$ and obtain
\begin{align}
\bar{{\cal{K}}}_{x=3}^{(y=2)}=\{k_{2}\},\quad {\cal{K}}_{x=3}^{(y=2)}=\{k_{1},k_{3},k_{4}\}\nonumber
\end{align}

\

$\bullet$ Step \pw{\ref{subyes?}} : answer is YES

\

The third subgraph of $\bar{{\cal{S}}}_{e}$ has $\{k_{2}\}$ as its internal momentum.

\

$\bullet$ Step \pw{\ref{divv}} : answer is YES

\

The divergent terms are 
\begin{align}
\int\limits_{k_{1}\cdots k_{4}}
&\Delta^{2}(k_{1})\left[\sum_{l=0}^{4}f_{l}^{\;(k_{1},\;p)}+{\bar{f}}^{\;(k_{1},\;p)}\!\right]^{2}\Delta(k_{2})\times\nonumber\\&\Delta(k_{2}\!-\!k_{1})\Delta(k_{2}-k_{3})\Delta(k_{3})\Delta(k_{3}-k_{1})\times\nonumber\\&\Delta(k_{4})\Delta(k_{4}-k_{1}+p).
\label{diver3}
\end{align}

$\bullet$ Step \pw{\ref{divok?}} : answer is YES

\

One may notice that (\ref{diver3}) contains the same terms of (\ref{diver}) and (\ref{diver1}).

\

$\bullet$ Step \pw{\ref{Ksub?}} : answer is YES

\

One may notice that $\bar{{\cal{K}}}_{x=3}^{(y=2)}$ is a subset of $\bar{{\cal{K}}}_{x=2}^{(y=1)}$.

\

$\bullet$ Step \pw{\ref{choose1}} : answer is YES

\

We choose $k_{2},k_{3},k_{4}$ and obtain
\begin{align}
\bar{{\cal{K}}}_{x=3}^{(y=2)}=\{k_{1}\},\quad {\cal{K}}_{x=3}^{(y=2)}=\{k_{2},k_{3},k_{4}\}\nonumber
\end{align}

\

$\bullet$ Step \pw{\ref{subyes?}} : answer is YES

\

None subgraph has $\{k_{1}\}$ as its internal momentum.

\

$\bullet$ Step \pw{\ref{choose1}} : answer is NO

\

$\bullet$ Step \pw{\ref{choose2}} : answer is YES

\

We choose $k_{1},k_{2},k_{3},k_{4}$ and obtain
\begin{align}
\bar{{\cal{K}}}_{x=4}^{(y=1)}=\emptyset,\quad {\cal{K}}_{x=4}^{(y=1)}=\{k_{1},k_{2},k_{3},k_{4}\}\nonumber
\end{align}

\

$\bullet$ Step \pw{\ref{subyes?}} : answer is NO

\

$\bullet$ Step \pw{\ref{choose1}} : answer is NO

\

$\bullet$ Step \pw{\ref{choose2}} : answer is NO

\

At this point, we write down the results stored:

\begin{enumerate}
\item $\bar{{\cal{K}}}_{x=2}^{(y=1)}=\{k_{2},k_{3}\}$ and eq. (\ref{diver})
\label{A}
\begin{align}
A_{1}^{(4)}\!\equiv\!\!\!\!\!&\int\limits_{k_{1}\cdots k_{4}}\!\!\!\!
\Delta^{2}(k_{1})\Delta^{2}(k_{1}-p)\Delta(k_{4})\Delta(k_{4}-k_{1}+p)\times\nonumber\\&\times\Delta(k_{2})\Delta(k_{2}\!-\!k_{1})\Delta(k_{2}\!-\!k_{3})\Delta(k_{3})\Delta(k_{3}\!-\!k_{1})\nonumber,
\end{align}
\item $\bar{{\cal{K}}}_{x=3}^{(y=1)}=\{k_{4}\}$ and eq. (\ref{diver1})
\label{B}
\begin{align}
A_{2}^{(4)}\!\equiv\!\!\!\!\!&\int\limits_{k_{1}\cdots k_{4}}\!\!\!\!
\Delta^{2}(k_{1})\Delta^{2}(k_{1}-p)\Delta(k_{4})\Delta(k_{4}-k_{1}+p)\times\nonumber\\&\times\Delta(k_{2})\Delta(k_{2}\!-\!k_{1})\Delta(k_{2}\!-\!k_{3})\Delta(k_{3})\Delta(k_{3}\!-\!k_{1})\nonumber,
\end{align}
\item Term counted twice (intersection between the cases (\ref{A}) and (\ref{B}))
\begin{align}
A_{3}^{(4)}\!\equiv\!\!\!\!\!&\int\limits_{k_{1}\cdots k_{4}}\!\!\!\!
\Delta^{2}(k_{1})\Delta^{2}(k_{1}-p)\Delta(k_{4})\Delta(k_{4}-k_{1}+p)\times\nonumber\\&\times\Delta(k_{2})\Delta(k_{2}\!-\!k_{1})\Delta(k_{2}\!-\!k_{3})\Delta(k_{3})\Delta(k_{3}\!-\!k_{1})\nonumber,
\end{align}
\end{enumerate}

The term $A_{1}^{(4)}$ is divergent as two momenta ($k_{2},k_{3}$) go to infinity. Therefore, we identify the subgraph that contains these momenta and apply the algorithm on it. The result can be written using a graphic notation as below \cite{Note4}
\begin{align}
\frac{ig^{4}}{2}\!\!\int\limits_{k_{2}k_{3}}\Delta(k_{2})\Delta(k_{2}\!-\!k_{1})\Delta(k_{2}\!-\!k_{3})\Delta(k_{3})\Delta(k_{3}\!-\!k_{1})\nonumber\\=ig^{4}\Bigg\{\frac{k_{1}^2b_{6}}{6}\left[2I_{log}^{(2)}(\lambda^2)-\frac{13}{3}I_{log}(\lambda^2)+\mbox{finite}\right]+\nonumber\\\quad\quad+\sum\limits_{i=2}^{3}\quad\int\limits_{k_{i}}\Delta(k_{i})\Delta(k_{i}\!-\!k_{1})\left[\frac{I_{log}(\lambda^{2})}{2}\right]\Bigg\}\nonumber
\end{align}
\begin{figure}[ht]
\begin{flushleft}
$\quad\;$\includegraphics{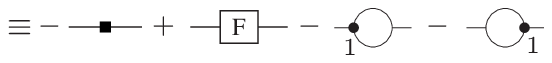}
\end{flushleft}
\end{figure}

Therefore, $A_{1}^{(4)}$ can be re-expressed as
\begin{figure}[h!]
\begin{center} 
\includegraphics{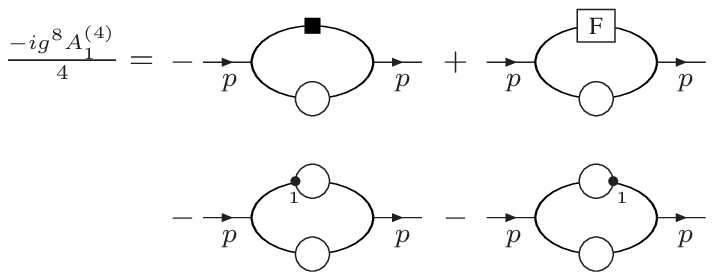}
\end{center}
\vspace{-0.50cm}
\end{figure}

We turn to the term $A_{2}^{(4)}$. Since it is divergent as only one momentum goes to infinity ($k_{4}$), we just apply the rules of IReg to obtain
\begin{align}
\frac{g^{2}}{2}\!\int\limits_{k_{4}}\!\!\!\Delta(k_{4})\Delta(k_{4}\!-\!p')\!=\!-\frac{g^2(p')^2}{6}\!\!\left[I_{log}(\lambda^2)+\mbox{finite}\right]\nonumber
\end{align}
\begin{figure}[h!]
\vspace{-0.9cm}
\begin{flushright} 
\includegraphics{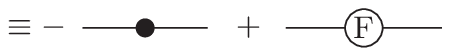}
\end{flushright}
\vspace{-0.4cm}
\end{figure}

\noindent
where $p'\equiv k_{1}-p$.

\vspace{0.15cm}
Thus, $A_{2}^{(4)}$ is given by
\begin{figure}[h!]
\begin{center}
\includegraphics{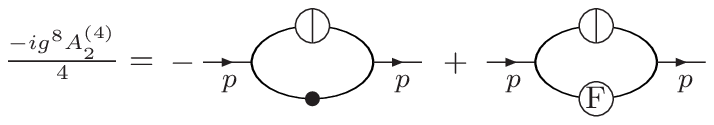}
\vspace{-0.7cm}
\end{center}
\end{figure}

We are left with the term $A_{3}^{(4)}$ which we can be re-expressed as

\begin{figure}[h!]
\begin{center} 
\includegraphics{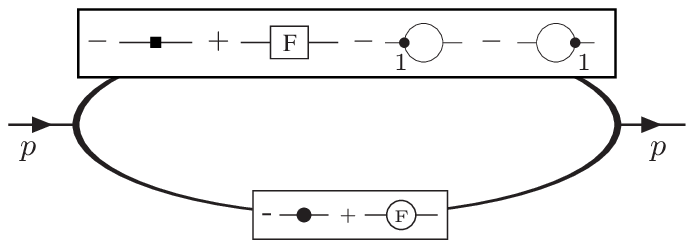}
\end{center}
\end{figure}

\noindent
or equivalently
\begin{figure}[h!]
\begin{center} 
\includegraphics{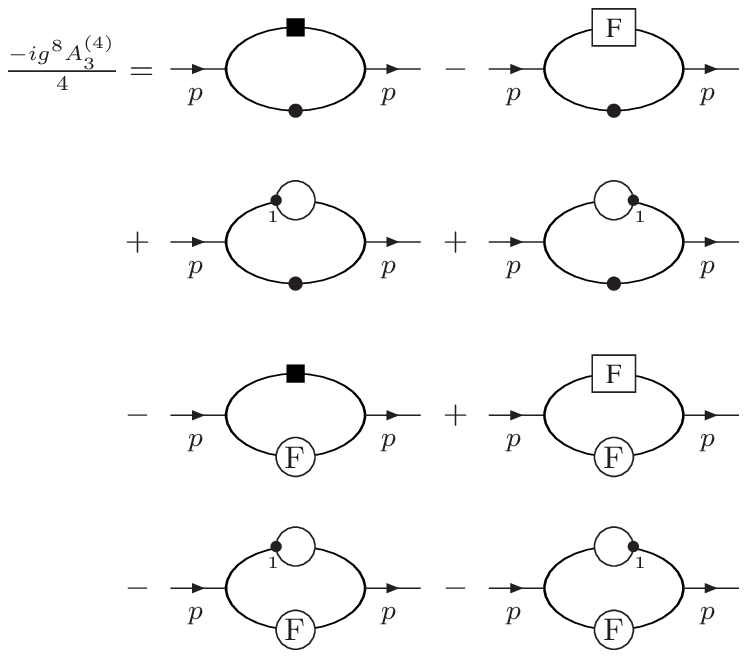}
\end{center}
\end{figure}

We set aside the terms that contain $I_{log}^{(l)}(\lambda^2)$ and run the algorithm again on the others. The results are:

\begin{figure}[h!]
\begin{flushleft} 
\includegraphics{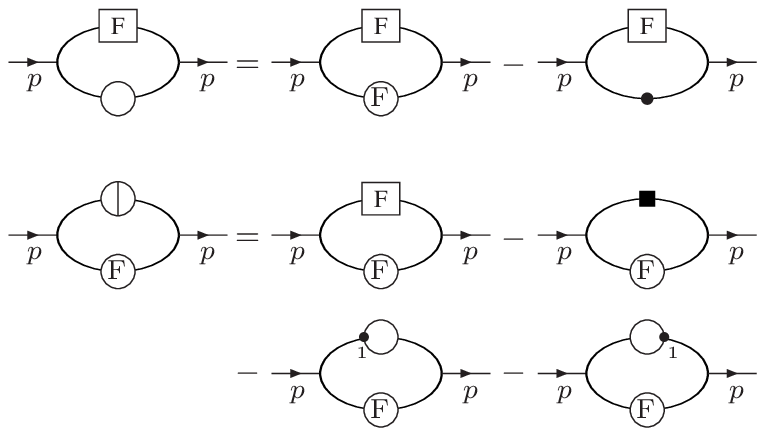}
\end{flushleft}
\end{figure}
\begin{figure}[h!]
\vspace{-0.8cm}
\begin{flushleft}
\includegraphics{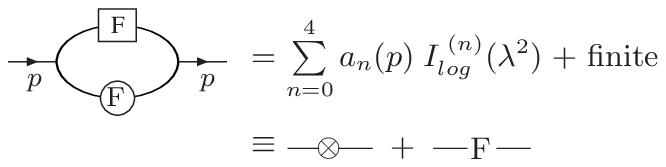}
\end{flushleft}
\end{figure}
%

A brief comment: although we have not calculated the finite part of $P_{A}^{(2)}$ (which appears in $A_{1}^{(4)}$), it can be shown to be of the form logarithm plus constants, allowing us to express all divergences in terms of $I_{log}^{\;(l)}(\lambda^2)$. 

At this point, we notice that all divergent terms contain $I_{log}^{\;(l)}(\lambda^2)$ and the divergent content of $\Xi_{A}^{(4)}$ is given by
\begin{align}
\Xi_{A}^{(4)\infty}=-\frac{ig^{8}}{4}\left[A_{1}^{(4)}+A_{2}^{(4)}-A_{3}^{(4)}\right]\quad\quad\quad\quad\quad\quad\nonumber
\end{align}
\begin{figure}[h!]
\vspace{-0.6cm}
\begin{center}
$\quad\quad\;\;\;$\includegraphics{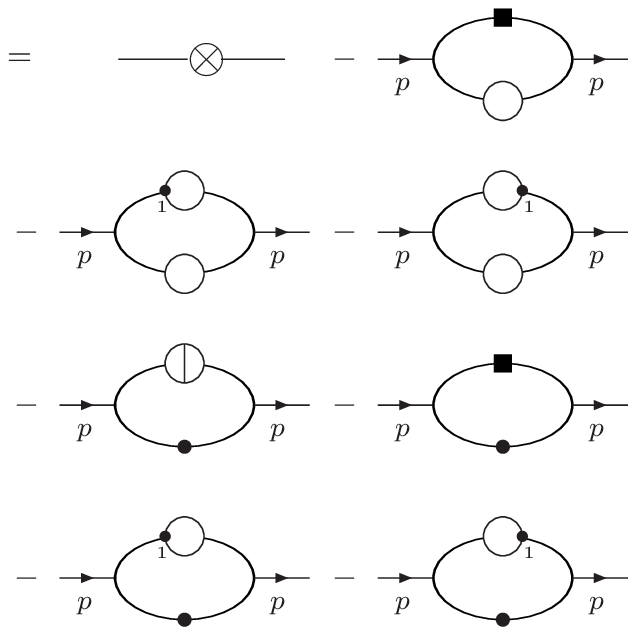}
\end{center}
\end{figure}
\vspace{-0.2cm}

As stated before, the first term (in which $I_{log}^{(l)}(\lambda^2)$ multiplies only constants and/or polynomials in the external momentum) corresponds to the typical divergence of the graph while the others (in which $I_{log}^{(l)}(\lambda^2)$ multiplies an integral) correspond to the terms to be subtracted by Bogoliubov's recursion formula.

\subsection{$N$-loop vertex and self-energy diagrams of the ladder type}

To conclude this section, we focus on $n$-loop graphs. We applied our method in two examples: the first contained only nested divergent subdiagrams while the other contained all kinds of subdivergences including overlapping ones. In both cases, our method found success since it was able to display the terms to be subtracted by Bogoliubov's recursion formula and the typical divergence of the graph was also written in terms of a well defined set of basic divergent integrals in one loop momentum only. Due to the lack of space, we just present the result of the first graph while the other one is discussed in more detail. The nested diagram reads
\begin{figure}[ht]
\begin{center} 
\includegraphics{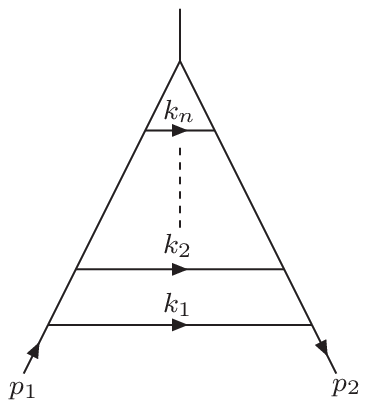}
\end{center}
\caption{$V^{(n)}$} 
\end{figure}

\noindent
and its amplitude, after the subtraction of the subdivergences, is given by
\begin{align}
\frac{\bar{\Lambda}^{(n)}}{i^{\;n+1}g^{2n+1}}\equiv \sum\limits_{m=1}^{n}c_{m}^{\;(n-1)}I_{log}^{(m)}(\lambda^2)+\mbox{finite},
\label{A6}
\end{align}
\noindent
where the coefficients $c_{m}^{\;(n-1)}$ can be recursively obtained
\begin{align}
c_{m=1}^{\;(0)}&\equiv1,\nonumber\\
c_{m=1}^{\;(n-1)}&\equiv2b_{6}\sum\limits_{p=1}^{n-1}\left(\frac{F_{p}}{p}+\frac{D_{p}}{p}\right)c_{p}^{\;(n-2)},\nonumber\\
c_{m\neq1}^{\;(n-1)}&\equiv2b_{6}\!\!\!\sum\limits_{p=m-1}^{n-1}\binom{p}{\;p-m+1}\frac{F_{p-m+1}}{p}c_{p}^{\;(n-2)},\nonumber\\
D_{q}&\equiv\frac{d^{q}}{d\epsilon^{q}}\Bigg[\frac{1}{(2-\epsilon)(1-\epsilon)}\Bigg]\Bigg|_{\epsilon=0},\nonumber\\
F_{q}&\equiv\frac{d^{q}}{d\epsilon^{q}}\Bigg[\frac{1}{(\epsilon+2)(\epsilon+1)(\epsilon-1)}\Bigg]\Bigg|_{\epsilon=0}.
\label{proof}
\end{align}

Before we proceed to the other graph, we state an important result which was essential to show that the typical divergence of $V^{(n)}$ is expressed in terms of BDI's (the procedure to compute finite integrals can be found in \cite{Dias:2010})
\begin{align}
&\int\limits_{k_{j}}\!\!\Delta(k_{j}\!-\!k_{a})\Delta^{2}(k_{j})\!\!\left[\!\sum\limits_{m=1}^{i}c_{m}^{\;(i-1)}\!\ln^{m-1}\!\!\left(\!\!-\frac{k_{j}^{2}\!-\!\mu^2}{\lambda^2}\!\right)\!\!\right]\!\!=\nonumber\\&\!\!\!\sum\limits_{m=1}^{i}c_{m}^{\;(i-1)}I_{log}^{(m)}(\lambda^2)+\sum\limits_{m=1}^{i+1}c_{m}^{\;(i)}\!\ln^{m-1}\!\left(\!\!-\frac{k_{a}^{2}-\!\mu^2}{\lambda^2}\right)
\label{integralkj}
\end{align}

Now we are ready to evaluate the graph with overlapped subdivergences which reads
\begin{figure}[h!]
\vspace{-0.25cm}
\begin{center} 
\includegraphics{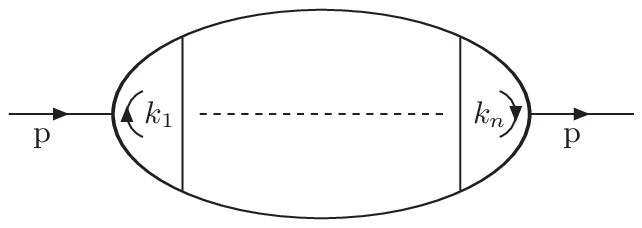}
\end{center}
\vspace{-0.45cm}
\caption{Graph $P^{(n)}$}
\end{figure}
\vspace{-0.25cm}

Its amplitude is given by
\begin{align}
&\Xi^{(n)}\!=\;{\cal{A}}\!\!\int\limits_{k_{1}\cdots k_{n}}\prod\limits_{r=1}^{n}\Delta(k_{r})\Delta(k_{r}\!-\!p)\!\prod\limits_{l=2}^{n}\Delta(k_{l}\!-\!k_{l-1}),
\\&\mbox{where}\quad{\cal{A}}\equiv\frac{i^{\;n-1}g^{2n}}{2}.\quad\quad\quad\quad\quad\quad\quad\quad\quad\quad\quad\quad\quad\quad\quad\quad\quad\quad\quad\quad\quad\quad\quad\quad\quad\quad\quad\quad\quad\quad\quad\quad\quad\quad\quad\nonumber
\end{align}

Before applying our method, we introduce a graphic notation suitable for our purposess

\begin{figure}[h!]
\vspace{-0.2cm}
\begin{center}
\includegraphics{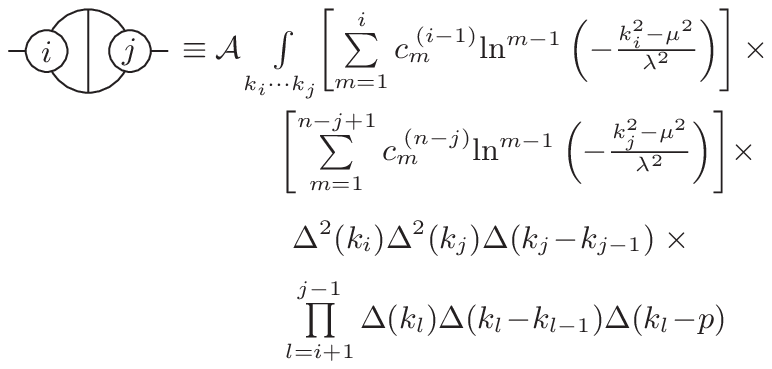}
\end{center}
\end{figure}
\vspace{-0.6cm}
\begin{figure}[h!]
\begin{center} 
\includegraphics{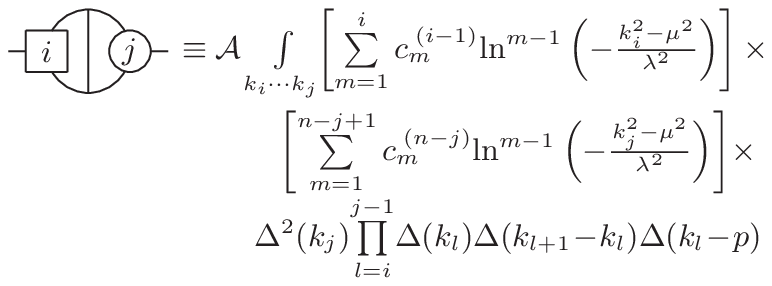}
\end{center}
\end{figure}

\begin{figure}[h!]
\begin{center}
\includegraphics{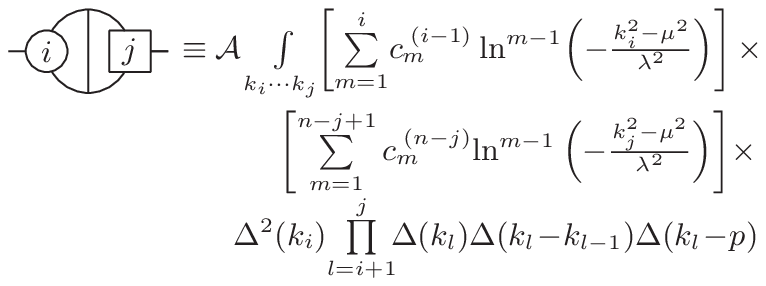}
\end{center}
\vspace{-0.2cm}
\end{figure}
\begin{figure}[h!]
\begin{center}
\includegraphics{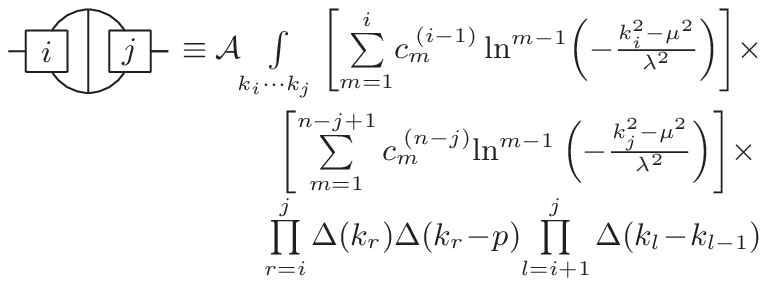}
\end{center}
\vspace{-0.2cm}
\end{figure}
\begin{figure}[h!]
\begin{center}
\includegraphics{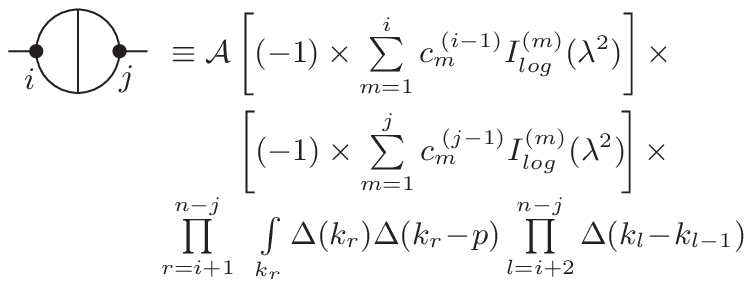}
\end{center}
\vspace{-0.2cm}
\end{figure}
\begin{figure}[h!]
\begin{center}
\includegraphics{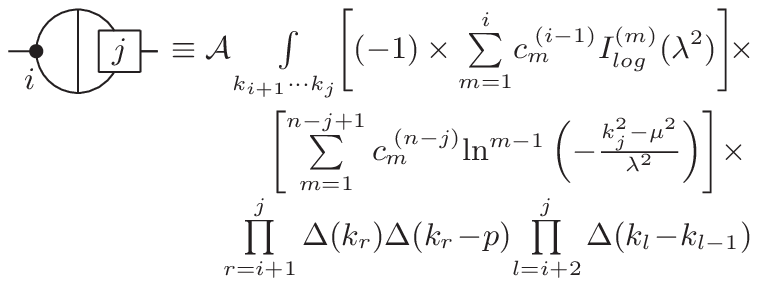}
\end{center}
\vspace{-0.2cm}
\end{figure}
\begin{figure}[h!]
\begin{center} 
\includegraphics{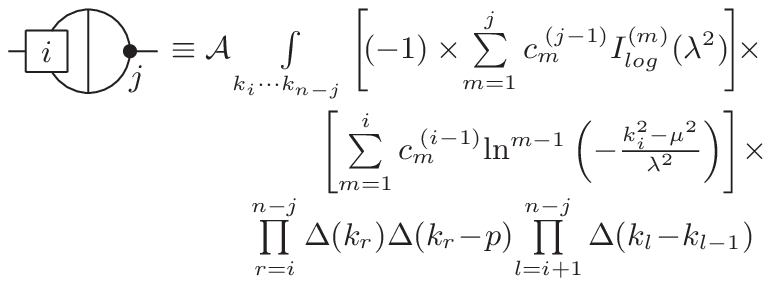}
\end{center}
\vspace{-0.35cm}
\end{figure}
\noindent
and define
\begin{align}
I_{\;i,\;j}\equiv\!\!\!\!\int\limits_{k_{i}\cdots k_{j}}\!\!\!&\left[\sum\limits_{m=1}^{i}\!c_{m}^{\;(i-1)}\!\ln^{m-1}\!\!\left(\!-\frac{k_{i}^{2}\!-\!\mu^2}{\lambda^2}\right)\!\!\right]\!\!\Delta^3(k_{i})\times\nonumber\\&\left[\sum\limits_{m=1}^{n-j+1}\!\!c_{m}^{\;(n-j)}\!\ln^{m-1}\!\!\left(\!-\frac{k_{j}^{2}\!-\!\mu^2}{\lambda^2}\right)\!\!\right]\!\!\Delta^3(k_{j})\times\nonumber\\&\frac{(2p\cdot k_{i})(2p\cdot k_{j})}{(k_{i+1}\!-\!k_{i})^2-\mu^2}\!\prod\limits_{l=i+1}^{j-1}\!\!\!\Delta^2(k_{l})\Delta(k_{l+1}\!-\!k_{l}).
\end{align}

In all the terms defined above, we consider that a bad-defined product is valued to unity. We are now in a position to use our algorithm but, since all divergent subgraphs share one leg with one of the external legs of the entire graph, it reduces to steps \ref{pro}-\ref{save}. Therefore, we just apply identity (\ref{identbphz}) in the propagators that depend on the external momenta, choose $n^{(k_{i})}=3$ ($i=1\cdots n$) and find the following divergent terms:

\begin{enumerate}
\item Divergences as $k_{1}\rightarrow \infty$ and $k_{2}\cdots k_{n}$ are fixed
\label{casekn1}
\begin{figure}[h!]
\begin{flushright}
\includegraphics{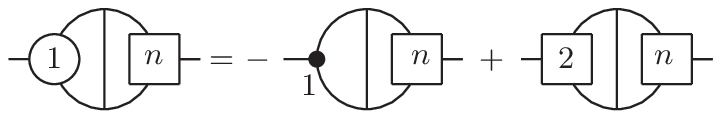}
\end{flushright}
\vspace{-0.1cm}
\end{figure}
\item Divergences as $k_{n}\rightarrow \infty$ and $k_{1}\cdots k_{n-1}$ are fixed
\label{casek1n}
\begin{figure}[h!]
\begin{flushright}
\includegraphics{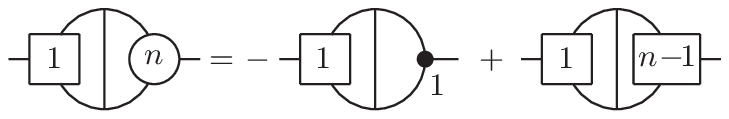}
\end{flushright}
\vspace{-0.1cm}
\end{figure}
\item Term counted twice (intersection between the cases (\ref{casekn1}) and (\ref{casek1n}))
\vspace{2.0cm}
\begin{figure}[h!]
\begin{flushright}
\includegraphics{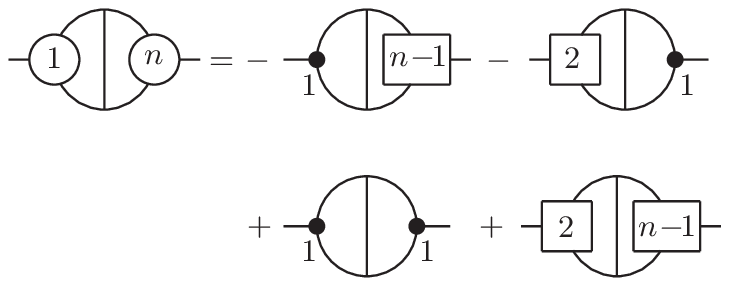}
\end{flushright}
\vspace{-0.5cm}
\end{figure}
\item Divergences as $k_{1}\cdots k_{n}\rightarrow \infty$ simultaneously
\begin{align}
I_{\;1,\;n}\nonumber
\end{align}
\end{enumerate}

At this point, we would set aside the terms that contain $I_{log}(\lambda^{2})$ and use the method again in the others. Although this is the basic procedure to be followed, adopting it will prove to be a harder way to find the typical divergence of the graph for arbitrary $n$. Therefore, instead of applying our method in a blind way we may notice that the results above were obtained after the use of (\ref{integralkj}) which means that we performed some of the $n$ integrals. In the first two cases we performed just one integral ($k_{1}$ and $k_{n}$ respectively), while in the third one we performed the integrals in $k_{1}$ and $k_{n}$. Hence, if we had performed only one integration, the divergent content of $P^{(n)}$ would be given by

\begin{figure}[h!]
\begin{center}
\includegraphics{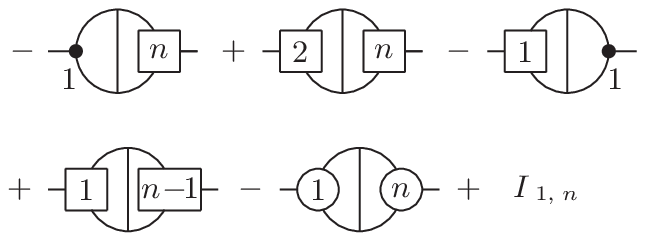}
\end{center}
\end{figure}
\vspace{-0.1cm}

Now, we suppose that performing $i-1$ integrals lead us to
\vspace{0.1cm}

\begin{figure}[h!]
\begin{center}
\includegraphics{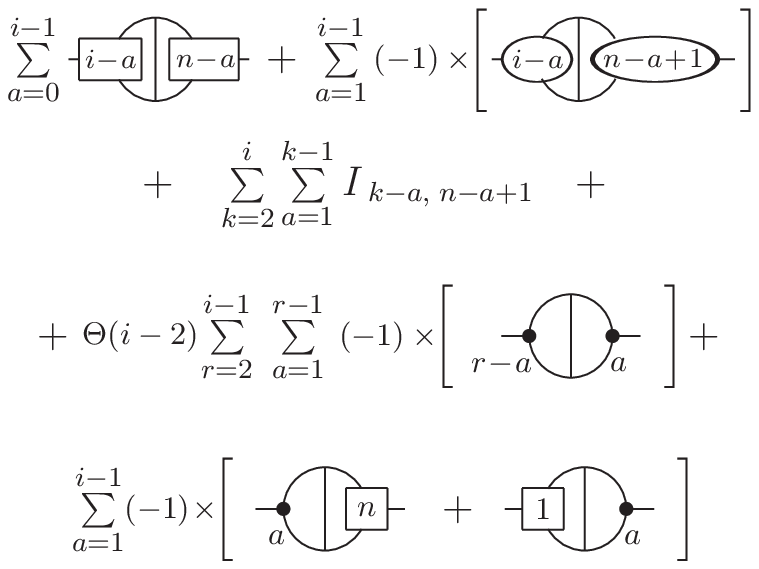}
\end{center}
\caption{}
\label{inductionp}
\vspace{-0.2cm}
\end{figure}

To prove that the above statement is correct, we use mathematical induction. We begin by applying our method to a typical term of the first summation and obtain the following divergent terms:

\begin{enumerate}
\item Divergences as \small$k_{i-a}\rightarrow \infty$ \normalsize and \small$k_{i-a+1}\cdots k_{n-a}$ \normalsize are fixed
\label{kn1}
\begin{figure}[h!]
\begin{flushright}
\includegraphics{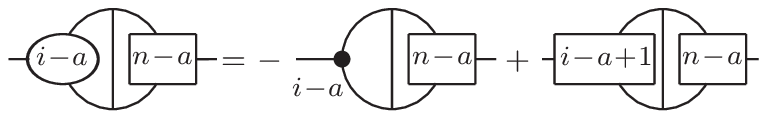}
\end{flushright}
\vspace{-0.6cm}
\end{figure}
\item Divergences as \small$k_{n-a}\rightarrow \infty$ \normalsize and \small$k_{i-a}\cdots k_{n-a-1}$ \normalsize are fixed
\label{k1n}
\begin{figure}[h!]
\begin{flushright}
\includegraphics{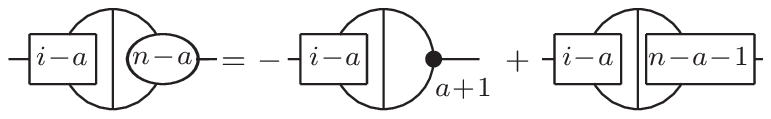}
\end{flushright}
\vspace{-0.6cm}
\end{figure}
\item Term counted twice (intersection between the cases (\ref{kn1}) and (\ref{k1n}))
\begin{figure}[h!]
\begin{center}
\includegraphics{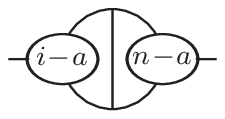}
\end{center}
\vspace{-0.6cm}
\end{figure}\item Divergences as \small$k_{i-a}\cdots k_{n-a}\rightarrow \infty$ \normalsize simultaneously
\begin{align}
I_{\;i-a,\;n-a}\nonumber
\end{align}
\end{enumerate}

In other words, the divergent content of a typical term of the first summation is given by

\begin{figure}[h!]
\begin{center}
\includegraphics{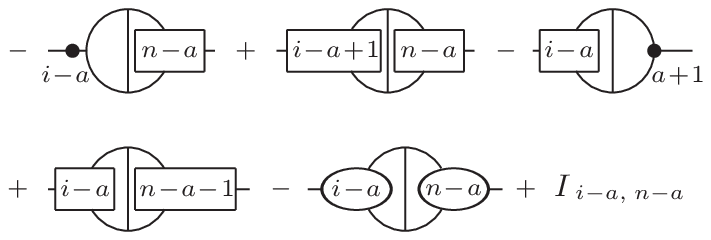}
\end{center}
\vspace{-0.5cm}
\caption{}
\label{plus}
\end{figure}
\vspace{-0.3cm}
 
We consider now a typical term contained in the second summation. As we seek the divergent content of $P^{(n)}$ after $i$ integrals are performed, we are able to re-express it as below

\begin{figure}[h!]
\begin{center}
\includegraphics{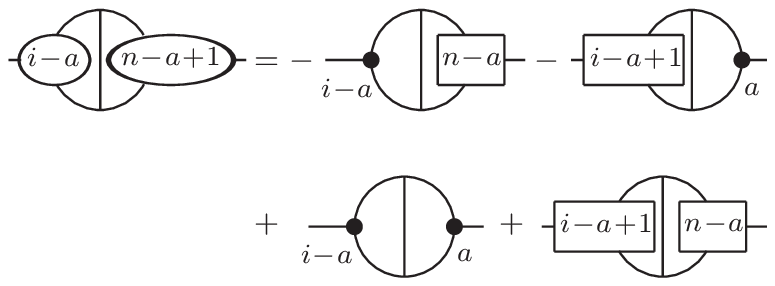}
\end{center}
\vspace{-0.5cm}
\caption{}
\label{minus}
\end{figure} 

Inserting the results of (fig \ref{plus}) and (fig \ref{minus}) at (fig \ref{inductionp}) we obtain

\vspace{1cm}
\begin{figure}[h!]
\vspace{-1cm}
\begin{center}
\includegraphics{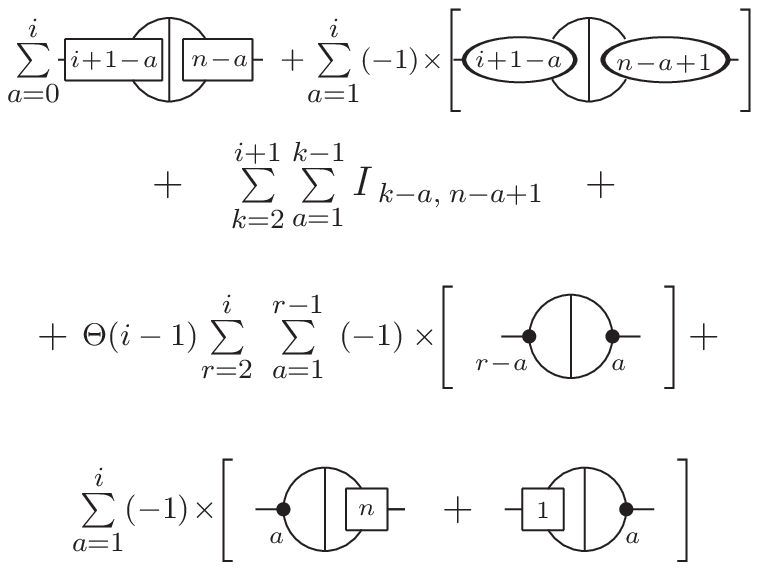}
\end{center}
\end{figure}

One may notice that the result above is just the one we supposed (fig \ref{inductionp}) if we set $i\rightarrow i+1$. Therefore, by mathematical induction, we prove our statement and we will be able to identify the typical divergence of the graph. Performing $n-1$ integrals lead us to

\begin{figure}[h!]
\begin{center}
\includegraphics{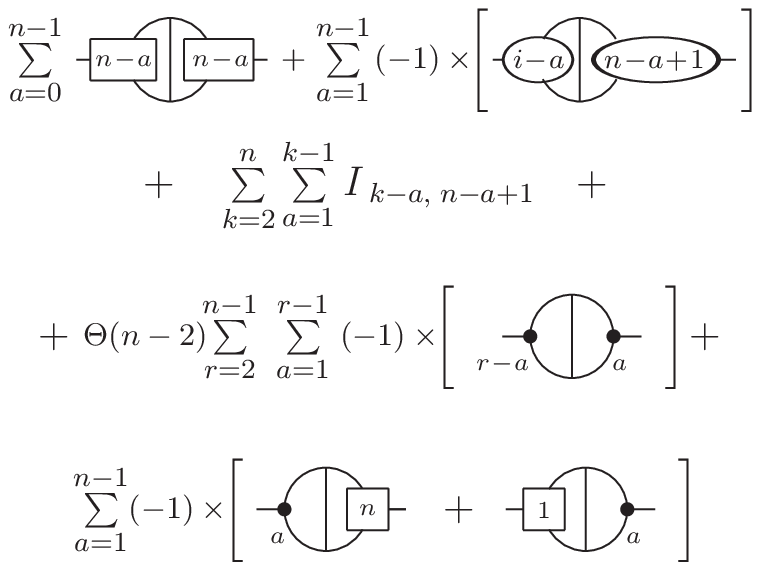}
\end{center}
\end{figure}

We apply our method in the first summation and find that the divergent terms coming from it are given by
\begin{align}
&p^2\sum\limits_{a=0}^{n-1}\sum\limits_{m=1}^{n-a}\sum\limits_{m'=1}^{a+1}c_{m}^{\;(n-a-1)}c_{m'}^{\;(a)}\left[-\frac{I_{log}^{\;(M-1)}(\lambda^2)}{3}+\right.\nonumber\\&\left.2\;\Theta(M-2)\!\sum\limits_{k=2}^{M-1}\left(\frac{1}{3}\right)^{k}\!\!\frac{(M-2)!}{(M-1-k)!}\;I_{log}^{\;(M-k)}(\lambda^2)\!\right]\!\!,\nonumber\\
&M\equiv m+m'.
\label{diag}
\end{align}

In the second summation all the terms contain only quadratic divergences and, therefore, they give a null contribution. The third summation can be evaluated and the result is
\begin{align}
&p^2\sum\limits_{k=2}^{n}\sum\limits_{a=1}^{k-1}\sum\limits_{m=1}^{k-1}\sum\limits_{m'=1}^{a}c_{m}^{\;(k-2)}d_{m'}^{\;(n-a-k+2)}\times\nonumber\\
&\left[2\!\!\sum\limits_{l=1}^{m+m'-1}\!\!\!\left(\frac{1}{3}\right)^{l}\!\!\frac{(m+m'-2)!}{(m+m'-1-l)!}I_{log}^{\;(m+m'-l)}(\lambda^2)\right],\nonumber\\
&d_{q}^{\;(0)}\equiv c_{k}^{(n-j)},\nonumber\\
&d_{q}^{\;(i)}\equiv 2b_{6}\sum\limits_{k=q}^{n-j+1}d_{k}^{\;(i-1)}\binom{k-1}{k-q}G_{k-q},\nonumber\\ 
&G_{q}\equiv\frac{d^{q}}{d(\epsilon a)^{q}}\left[\frac{1}{[4-(\epsilon a)^2][1-(\epsilon a)^2]}\right]\Bigg|_{\epsilon=0}.
\label{I's}
\end{align}

The others terms are just the ones that are subtracted by Bogoliubov's recursion formula since the counterterms for the graph we are dealing with can be re-expressed as

\begin{figure}[h!]
\begin{center}
\includegraphics{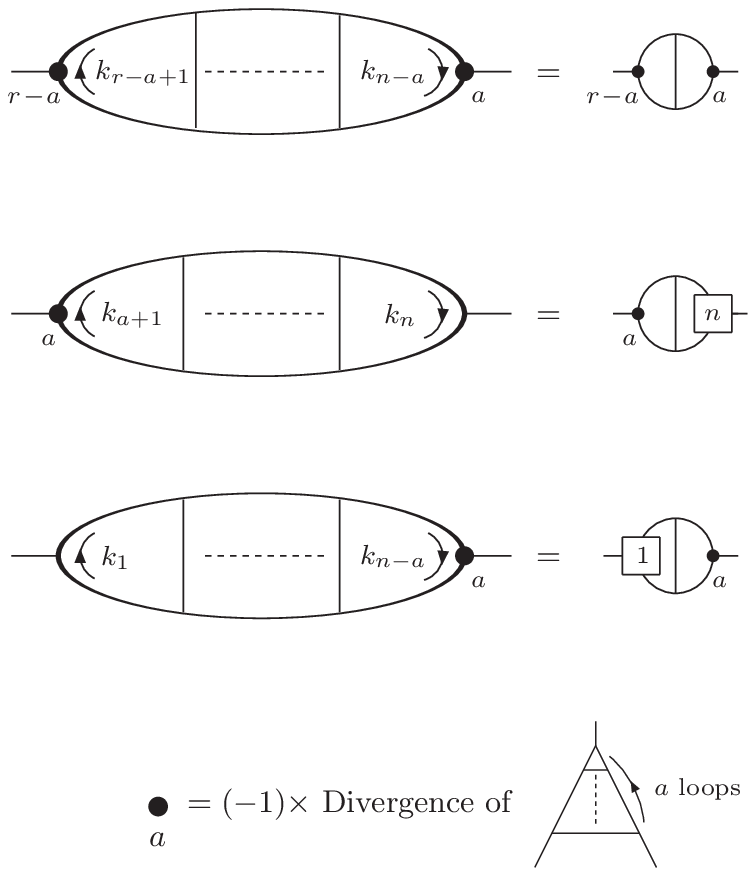}
\end{center}
\vspace{-0.3cm}
\end{figure}

Therefore, the typical divergence of $P^{(n)}$ (after the subtraction of subdivergences) is given by the sum of (\ref{diag}) with (\ref{I's}).

One may notice that our method once again displayed the terms to be subtracted by Bogoliubov's recursion formula and allowed us to express the typical divergence of the graph in terms of well defined BDI's. Therefore, we claim that the procedure we developed here (which is summarized in the algorithm) is the systematization of IReg to multi-loop Feynman graphs. Although the results we already obtained are restricted to massless theories, we can develop a similar procedure to treat massive ones as shown in the next section.

\section{Massive theories}
\label{s:mass}

We generalize identity (\ref{identbphz}) to massive theories and implement a mass independent regularization scheme. The procedure we develop here allows us to write the divergences of any theory (massive or not) in terms of the set of basic divergent integrals we have already presented and, therefore, justifies our restriction to massless theories in the previous sections.

The identity we are going to use is
\begin{align}
\frac{1}{(k\!-\!p_i)^2\!-m^{2}\!-\!\mu^{2}}=\!\!\!\!\!\sum_{l=0}^{2(n^{(k)}_{i}-1)}\!\!\!h_{l}^{(k,\;p_{i})}\!+{\bar{h}}^{\;(k,\;p_{i})}\!,
\label{identm}
\end{align}
\noindent
where we defined,
\begin{align}
&h_{l}^{(k,\;p_{i})}\!\equiv\sum_{j=0}^{\left\lfloor l/2\right\rfloor}\!\Theta(n^{(k)}_{i}+j-l)\binom{l-j}{j}\times\nonumber\\&\qquad\qquad\qquad\qquad\times\frac{(-p_i^2+m^2)^{j}(2p_i \cdot k)^{l-2j}}{(k^2-\mu^2)^{l+1-j}}\\
&{\bar{h}}^{\;(k,\;p_{i})}
\equiv\frac{(-1)^{n^{(k)}_{i}}\!(p_i^2\!-\!m^2-2p_i \cdot k)^{n^{(k)}_{i}}}{(k^2\!-\!\mu^2)^{n^{(k)}_{i}}
\!\left[(k\!-\!p_i)^2\!-m^2\!-\!\mu^{2}\right]}\!.
\end{align}

Notice that, if we set $m^{2}\rightarrow 0$, we recover identity (\ref{identbphz}). The procedure to be followed is similar to the one of the massless theory since $h_{l}^{(k,\;p_{i})}$ was also constructed in such a way that it goes like $k^{-(l+2)}$ as $k\rightarrow \infty$. The main difference is that we must apply identity (\ref{identm}) in all propagators which depend on the external momenta and/or the mass. In order to familiarize the reader with this identity, we evaluate the one-loop contribution for the propagator 

\begin{figure}[ht]
\begin{center}
\includegraphics{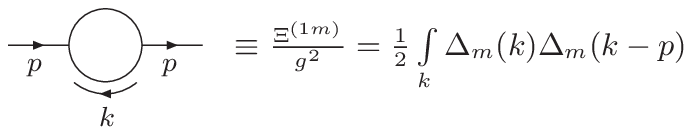}
\end{center}
\vspace{-0.2cm}
\caption{Graph $M^{(1)}$}
\end{figure}
 
In the above picture we defined \cite{Note3}
\begin{align}
\Delta_{m}(k)\equiv\frac{1}{k^2-m^2-\mu^{2}}.
\end{align}

We apply (\ref{identm}) in all denominators to obtain (we defined $p_{0}=0$ and $p_{1}=p$)
\begin{align}
\frac{\Xi^{(1m)}}{g^2}=\frac{1}{2}\int\limits_{k}&\left[\sum_{l=0}^{2(n^{(k)}_{0}-1)}h_{l}^{(k,\;0)}+{\bar{h}}^{\;(k,\;0)}\!\right]\times\nonumber\\&\times\left[\!\sum_{q=0}^{2(n^{(k)}_{1}-1)}h_{l}^{(k,\;p)}+{\bar{h}}^{\;(k,\;p)}\!\right]\!\!.
\end{align}

As usual, we choose the minimum value of $n^{(k)}_{i}$ ($i=0,1$) in order to guarantee the finitude of the terms that contain ${\bar{h}}^{\;(k,\;p_{i})}$ as $k\rightarrow \infty$. In this case, we find that $n^{(k)}_{0}=n^{(k)}_{1}=3$. At this point we identify the divergent terms:

\begin{enumerate}
\item Quadratic divergence
\begin{align}
\int\limits_{k}h_{0}^{(k,\;0)}h_{0}^{(k,\;p)}=\int\limits_{k}\frac{1}{(k^{2}-\mu^{2})^{2}},
\end{align}
\item Linear divergence
\begin{align}
\int\limits_{k}h_{0}^{(k,\;0)}h_{1}^{(k,\;p)}=\int\limits_{k}\frac{2p\cdot k}{(k^{2}-\mu^{2})^{3}},
\end{align}
\item Logarithmic divergence
\begin{align}
\int\limits_{k}\!h_{2}^{(k,\;0)}h_{0}^{(k,\;p)}=\int\limits_{k}\frac{m^{2}}{(k^{2}-\mu^{2})^{3}},
\end{align}
\begin{align}
\!\!\!\int\limits_{k}\!h_{0}^{(k,\;0)}h_{2}^{(k,\;p)}\!=\!\!\int\limits_{k}\!\left[\frac{(2p\cdot k)^{2}}{(k^{2}\!-\!\mu^{2})^{4}}-\frac{(p^2\!-m^2)}{(k^{2}\!-\!\mu^{2})^{3}}\!\right]\!\!.
\end{align}
\end{enumerate}

One may notice that the quadratic and linear divergences are the same of the massless case and they give null contributions. Therefore, after the use of the rules of IReg, the remaining terms yield
\begin{align}
\frac{\Xi^{(1m)}}{g^{2}}=-\!\!\left[\!\left(\frac{p^2}{6}-m^2\right)\!I_{log}(\lambda^{2})+\frac{p^2}{3}\Upsilon_{1}+\mbox{finite}\right]\!\!.
\label{prop1}
\end{align}

The two-loop corrections are obtained in the same way and we just present the results. After the subtraction of the subdivergences, the two-loop overlapped graph ($P_{A}^{(2)}$) furnishes
\begin{align}
\frac{\bar{\Xi}^{(2m)}_{A}}{ig^4}\!=&\!\left[\!\frac{I_{log}^{(2)}(\lambda^2)}{3}-\frac{13}{18}I_{log}(\lambda^2)+9\Upsilon_{1}-4\Upsilon_{2}\right]\!b_{6}p^{2}\nonumber\\&-\Bigg[2I_{log}^{(2)}(\lambda^2)+5I_{log}(\lambda^2)\!\Bigg]\!b_{6}m^2+\mbox{finite},
\end{align}
\noindent
while for the two-loop nested graph ($P_{B}^{(2)}$) we obtain
\begin{align}
\frac{\bar{\Xi}_{B}^{(2m)}\!\!\!\!}{ig^{4}}\!&=\!\Bigg\{\Bigg(-\frac{b_{6}p^2}{18}-\frac{b_{6}m^2}{2}\Bigg)I_{log}^{(2)}(\lambda^2)+\nonumber\\
&\!\!\!+\!\!\Bigg[\!\Bigg(\!\frac{5b_{6}}{27}-\!\frac{2\Upsilon_{1}}{3}\!\!\Bigg)p^{2}\!+\Bigg(\!\frac{2b_{6}}{3}+\!6\Upsilon_{1}\!\!\Bigg)m^{2}\!\Bigg]\!I_{log}(\lambda^2)\nonumber\\&\!\!\!+\!\!\Bigg(\frac{2b_{6}\Upsilon_{2}}{3}-\frac{4b_{6}\Upsilon_{1}}{9}+2(\Upsilon_{1})^{2}\Bigg)p^{2}+\mbox{finite}\Bigg\}.
\end{align}

Therefore, the renormalization of the propagator at two-loop order is given by
\begin{align}
\frac{\bar{\Xi}^{(2m)}\!\!\!\!}{ig^{4}}\!&=\!\Bigg(\frac{5b_{6}p^2}{18}-\frac{5b_{6}m^2}{2}\Bigg)I_{log}^{(2)}(\lambda^2)-\nonumber\\
&\!\!\!-\!\!\Bigg[\!\Bigg(\!\frac{29b_{6}}{54}+\!\frac{2\Upsilon_{1}}{3}\!\!\Bigg)p^{2}\!-\Bigg(\!\frac{17b_{6}}{3}+\!6\Upsilon_{1}\!\!\Bigg)m^{2}\!\Bigg]\!I_{log}(\lambda^2)\nonumber\\&\!\!\!-\!\!\Bigg(\frac{10b_{6}\Upsilon_{2}}{3}-\frac{77b_{6}\Upsilon_{1}}{9}-2(\Upsilon_{1})^{2}\Bigg)p^{2}+\mbox{finite}.
\label{prop2}
\end{align}

To conclude, we calculate the renormalization group functions. As usual, we make the definitions
\be
\phi_{o}\equiv Z_{\phi}^{\frac{1}{2}}\phi, \quad m_{o}\equiv Z_{m}^{\frac{1}{2}}m, \quad g_{o}\equiv Z_{g}g,
\ee
\noindent
which allows us to divide the bare lagrangian in two parts: the renormalized lagrangian and the counterterms. Explicitly,
\begin{align}
&L=\frac{1}{2}\left[(\partial_{\mu}\phi)^{2}-m^{2}\phi^{2}\right]+\frac{g}{3!}\phi^{3}+\nonumber\\&\quad\quad+\frac{1}{2}\left[A(\partial_{\mu}\phi)^{2}-Bm^{2}\phi^{2}\right]+\frac{g}{3!}C\phi^{3},\nonumber\\
&A\equiv Z_{\phi}-1, \quad B\equiv Z_{\phi}Z_{m}-1, \quad C\equiv Z_{g}Z_{\phi}^{\frac{3}{2}}-1
\label{Z}.
\end{align}

We also define the renormalization group functions by
\begin{align}
\gamma\equiv \frac{\lambda}{2}\frac{\partial \ln Z_{\phi}}{\partial \lambda}, \quad \beta \equiv \lambda \frac{\partial g}{\partial \lambda} =-g \lambda \frac{\partial \ln Z_{g}}{\partial \lambda},\\
\gamma_{m}\equiv -\frac{2}{m}\lambda\frac{\partial m}{\partial \lambda}=\lambda\frac{\partial \ln Z_{m}}{\partial \lambda}.
\label{beta}
\end{align}

Supposing that $A$, $B$ and $C$ have a expansion in the coupling constant $g$ as below
\begin{align}
A&=A_{1}g^{2}+A_{2}g^{4}+O(g^{6}),\\
B&=B_{1}g^{2}+B_{2}g^{4}+O(g^{6}),\\
C&=C_{1}g^{2}+C_{2}g^{4}+O(g^{6}),
\end{align}
\noindent
we obtain
\begin{align}
\!\!\!\!\!\!\ln Z_{\phi}=&A_{1}g^{2}+\left(A_{2}-\frac{A_{1}^{2}}{2}\right)g^{4}+O(g^{6}),\\
\!\!\!\!\!\!\ln Z_{m}=&\left(B_{1}-A_{1}\right)g^{2}+\nonumber\\
&\!\left(B_{2}-A_{2}+\frac{A_{1}^{2}}{2}-\frac{B_{1}^{2}}{2}\right)\!g^{4}\!+\!O(g^{6}),\\
\!\!\!\!\!\!\ln Z_{g}=&\left(C_{1}-\frac{3A_{1}}{2}\right)g^{2}+\nonumber\\
&\!\left(\!C_{2}-\!\frac{3A_{2}}{2}+\!\frac{3A_{1}^{2}}{4}-\!\frac{C_{1}^{2}}{2}\right)\!g^{4}\!+\!O(g^{6}).
\end{align} 

Since $\gamma$, $\gamma_{m}$ and $\beta$ can also be written as
\begin{align}
\gamma&=\gamma_{1}g^{2}+\gamma_{2}g^{4}+O(g^{6}),\\
\gamma_{m}&=\gamma_{m}^{(1)}g^{2}+\gamma_{m}^{(2)}g^{4}+O(g^{6}),\\
\beta&=\beta_{1}g^{3}+\beta_{2}g^{5}+O(g^{6}),
\end{align}
\noindent
we deduce the relations below
\begin{align}
\gamma_{1}&=\frac{\lambda}{2}\frac{\partial A_{1}}{\partial \lambda},\\
\gamma_{2}&=\beta_{1}A_{1}+\frac{\lambda}{2}\frac{\partial }{\partial \lambda}\left(A_{2}-\frac{A_{1}^{2}}{2}\right),\\
\gamma_{m}^{(1)}&=\lambda\frac{\partial }{\partial \lambda}(B_{1}-A_{1}),\\
\gamma_{m}^{(2)}&=2\left(B_{1}-A_{1}\right)\beta_{1}+\lambda\frac{\partial D}{\partial \lambda},\\
\beta_{1}&=-\lambda\frac{\partial }{\partial \lambda}\left(C_{1}-\frac{3A_{1}}{2}\right),\\
\beta_{2}&=-2\left(C_{1}-\frac{3A_{1}}{2}\right)\beta_{1}-\lambda\frac{\partial E}{\partial \lambda},\\
&D\equiv\left(B_{2}-A_{2}+\frac{A_{1}^{2}}{2}-\frac{B_{1}^{2}}{2}\right),\\
&E\equiv\left(C_{2}-\frac{3A_{2}}{2}+\frac{3A_{1}^{2}}{4}-\frac{C_{1}^{2}}{2}\right).
\end{align}

Because we chose a ``MS" scheme in IReg, the coefficients $A_{i}$, $B_{i}$ and $C_{i}$ are given by (equations (\ref{prop1}), (\ref{prop2}), (\ref{vert}) and fig. \ref{v1})
\begin{align}
&\!\!\!\!A_{1}=-\frac{i}{6}I_{log}(\lambda^2),\\
&\!\!\!\!A_{2}=-\frac{5b_{6}}{18}I_{log}^{(2)}(\lambda^2)+\!\Bigg(\!\frac{29b_{6}}{54}+\!\frac{2\Upsilon_{1}}{3}\!\!\Bigg)\!I_{log}(\lambda^2),\\
&\!\!\!\!B_{1}=-iI_{log}(\lambda^2),\\
&\!\!\!\!B_{2}=-\frac{5b_{6}}{2}I_{log}^{(2)}(\lambda^2)+\!\Bigg(\!\frac{17b_{6}}{3}+6\Upsilon_{1}\!\Bigg)\!I_{log}(\lambda^2),\\
&\!\!\!\!C_{1}=-iI_{log}(\lambda^2),\\
&\!\!\!\!C_{2}=-\frac{5b_{6}}{2}I_{log}^{(2)}(\lambda^2)+\!\Bigg(\!\frac{17b_{6}}{3}+6\Upsilon_{1}\!\Bigg)\!I_{log}(\lambda^2).
\end{align}

Since we have the relations
\begin{align}
\lambda\frac{\partial I_{log}(\lambda^2)}{\partial \lambda}=2\lambda^2\frac{\partial I_{log}(\lambda^2) }{\partial \lambda^2}=-2b_{6},\\
\lambda\frac{\partial I_{log}^{(2)}(\lambda^2) }{\partial \lambda}=-2I_{log}(\lambda^2)-3b_{6},
\end{align} 
\noindent
we finally obtain
\begin{align}
\gamma&=\frac{g^2}{12(4\pi)^{3}}+\frac{13g^4}{432(4\pi)^{6}}+\frac{ig^{4}}{3(4\pi)^{3}}\Upsilon_{1}+O(g^{6}),\\
\gamma_{m}\!&=\frac{5g^2}{6(4\pi)^{3}}+\frac{97g^4}{108(4\pi)^{6}}+\frac{16ig^{4}}{3(4\pi)^{3}}\Upsilon_{1}+O(g^{6}),\\
\beta&=-\frac{3g^3}{4(4\pi)^{3}}-\frac{125g^5}{144(4\pi)^{6}}-\frac{5ig^{5}}{(4\pi)^{3}}\Upsilon_{1}\!+O(g^{6}).
\end{align}

One may notice the appearance of arbitrary surface terms in the renormalization group functions. At first glance, this feature may be a result of the definition of the ``MS" scheme in IReg which corresponds to the subtraction of basic divergent integrals only. However, if we use a different scheme in which we subtract basic divergent integrals and surface terms we obtain the same result above for the first two coefficients of the $\beta$ function. 

As pointed out earlier, surface terms are related to gauge and supersymmetry in such a manner that setting them to zero guarantees the invariance of the amplitude regarding these symmetries. In our present case we are not dealing with gauge or supersymmetric theories but with a scalar one. Therefore, it is natural to ask if the surface terms will play any role and the answer is that they introduce an arbitrariness in the first two coefficients of the $\beta$ function which are known to be universal. Hence, we must set them to zero even in a scalar theory. This in turn corresponds to assure momentum routing invariance in the loops of an arbitrary Feynman diagram \cite{BaetaScarpelli:2000zs} and allow us to to conjecture that momentum routing invariance is a fundamental symmetry of Feynman graphs.

It is also important to note that when anomalies come into play, they may express themselves, from the perturbative viewpoint, as an explicit dependence in momentum routing as pointed out by Jackiw \cite{Treiman:1986ep}. In such case, to exhibit democratically the anomaly among the Ward identities, surface terms should be left arbitrary \cite{Souza:2005vf}.

\section{Concluding Remarks}

We have shown that IReg being a strong candidate for a symmetry preserving invariant regularization is consistent to locality, Lorentz invariance, unitarity and causality. This was achieved through an algorithm that implements IReg to multi-loop Feynman graphs in such a way that the terms to be subtracted by Bogoliubov's recursion formula are displayed automatically.
We have also demonstrated that CIReg (which corresponds to setting the surface terms to zero and is the sufficient condition to deliver gauge and supersymmetric invariant Green' s functions) should be adopted in theories with less symmetry content as well. We learn from this that momentum routing invariance is a fundamental symmetry of any Feynman diagram and all regularization procedures should comply with it at the expense of bringing arbitrary non physical parameters into the amplitude. In our example this is manifest in the two-loop coefficient of the $\beta$ function. It is important to notice that IReg is a $n$-loop invariant program that displays the divergences as basic divergent integrals. If one evaluate the BDI's or surface terms using a specific regularization (dimensional, Pauli-Villars, etc) then it becomes apparent that any regularization which attributes a non zero value to surface terms could crash with momentum routing invariance and thus assign a non physical value to the universal coefficients of the $\beta$ function. An exceptional case corresponds to anomalies in perturbation theory which manifests themselves as a breaking of momentum routing invariance \cite{Treiman:1986ep}. Then, for instance, in order to democratically display the anomaly between the vector and axial sectors of the AVV triangle (ABJ anomaly) surface terms must be let arbitrary as physical free parameters. 

\section*{Acknowledgements}

This work was supported by CNPQ.

\end{document}